# Observation of dendrite formation at Li metal-electrolyte interface: A machine-learning enhanced constant potential framework

Taiping Hu[1,2], Haichao Huang[3], Guobing Zhou[1,4], Xinyan Wang[5], Jiaxin Zhu[6], Zheng Cheng[2,7], Fangjia Fu[2,7], Xiaoxu Wang[5], Fuzhi Dai[2,8], Kuang Yu[*,3], Shenzhen Xu[*,1,2]

[1]Beijing Key Laboratory of Theory and Technology for Advanced Battery Materials, School of Materials Science and Engineering, Peking University, Beijing 100871, People's Republic of China

[2]AI for Science Institute, Beijing 100084, People's Republic of China

[3]Tsinghua-Berkeley Shenzhen Institute and Institute of Materials Research (iMR), Tsinghua Shenzhen International Graduate School, Tsinghua University, Shenzhen, 518055, People's Republic of China

[4]School of Chemical Engineering, Jiangxi Normal University, Nanchang 330022, People's Republic of China

[5]DP Technology, Beijing 100080, People's Republic of China

[6]State Key Laboratory of Physical Chemistry of Solid Surfaces, iChEM, College of Chemistry and Chemical Engineering, Xiamen University, Xiamen 361005, People's Republic of China

[7]School of Mathematical Sciences, Peking University, Beijing 100871, People's Republic of China

[8]School of Materials Science and Engineering, University of Science and Technology Beijing, Beijing, 100083, People's Republic of China

[*]Corresponding authors: yu.kuang@sz.tsinghua.edu.cn, xushenzhen@pku.edu.cn

# TOC Graphic

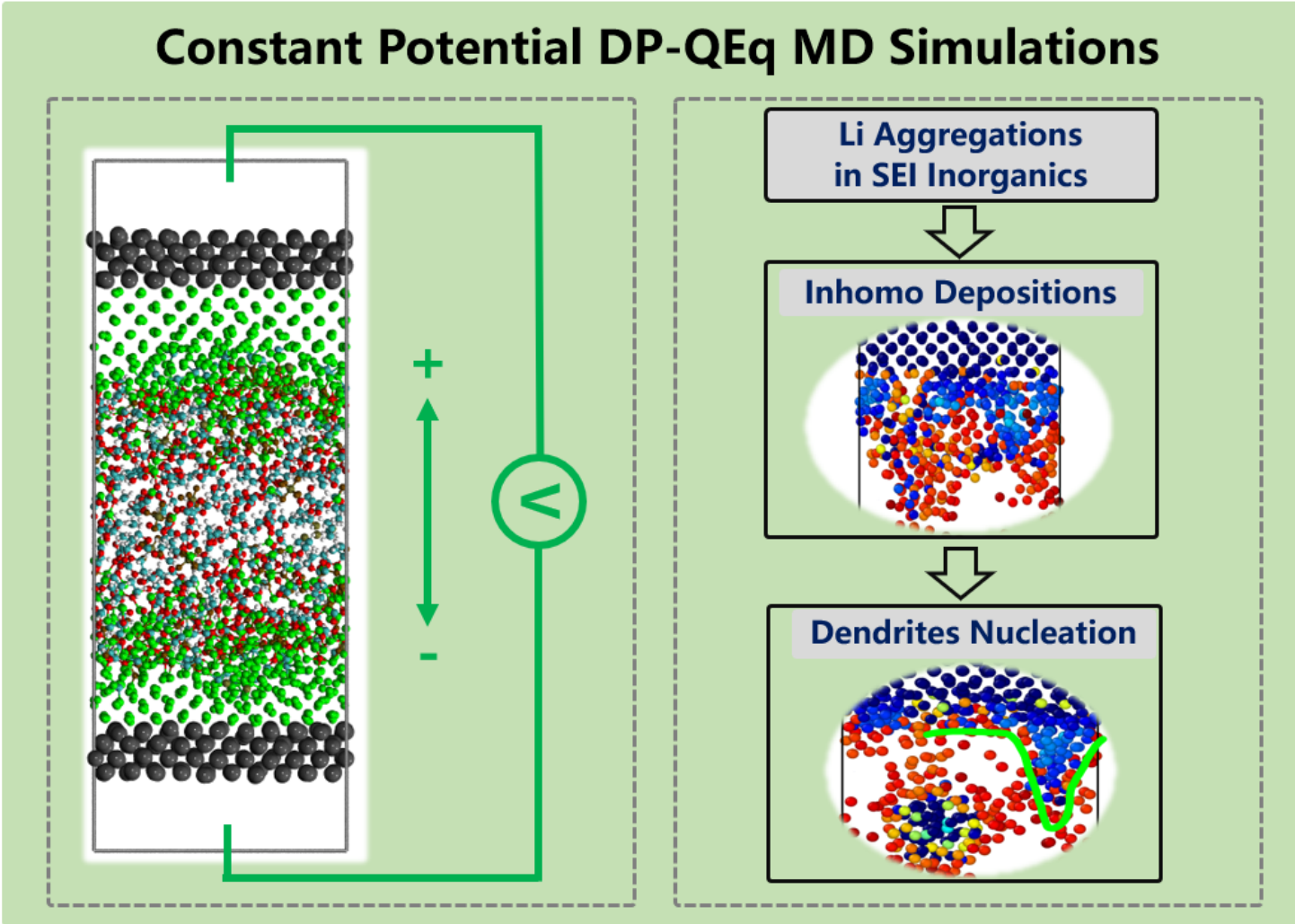

# Abstract

Uncontrollable dendrites growth during electrochemical cycles leads to low Coulombic efficiency and critical safety issues in Li metal batteries. Hence, a comprehensive understanding of the dendrite formation mechanism is essential for further enhancing the performance of Li metal batteries. Machine learning accelerated molecular dynamics (MD) simulations can provide atomic-scale resolution for various key processes at an *ab-initio* level accuracy. However, traditional MD simulation tools hardly capture Li electrochemical depositions, due to lack of an electrochemical constant potential (ConstP) condition. In this work, we propose a ConstP approach that combines a machine learning force field with the charge equilibration method to reveal the dynamic process of dendrites nucleation at Li metal anode surfaces. Our simulations show that inhomogeneous Li depositions, following Li aggregations in amorphous inorganic components of solid electrolyte interphases, can initiate dendrites nucleation. Our study provides microscopic insights for Li dendrites formations in Li metal anodes. More importantly, we present an efficient and accurate simulation method for modeling realistic ConstP conditions, which holds considerable potential for broader applications in modeling complex electrochemical interfaces.

## Introduction

Li-ion batteries (LIB) have received significant attention from both the industry and academia because of their extensive applications in electrochemical energy storage systems.[1,2,3,4] With an exceptionally high theoretical specific capacity (3860 mAh/g) and low density (0.59 g/cm$^3$), Li metal is considered as one of the most promising anodes for the next generation of high-energy-density batteries.[5,6,7,8] However, unexpected Li dendrite growths during electrochemical cycles lead to safety issues and poor battery performance, which severely impedes applications of Li metal anodes.[9,10,11] Effective suppression of the dendritic formation requires a deep understanding of its evolution dynamics. Experimentally, microscopic morphologies of Li dendrites can be observed by various advanced characterization methods, such as the scanning electron microscopy (SEM) and the transmission electron microscopy (TEM).[12,13,14] However, due to limited resolutions in both spatial and temporal scales, it is challenging for current characterization methods to capture and visualize the dynamics of dendrites formation.

Molecular dynamics (MD) simulations can provide atomic-scale insights into various complex processes. However, characterizing Li electro-deposition/dissolution through MD simulations presents a methodological challenge, due to the requirement of achieving an electrochemical contact condition, rather than a simple chemical contact, for modeling the electrode/electrolyte interface. Considering Li dendrites formation originates from charging/discharging cycles, constant potential (ConstP) MD simulation techniques are highly desired to describe those processes. In ConstP MD simulations, the atomic charges at electrochemical interfaces are dynamically updated to maintain applied potentials on electrodes, thus enabling us to model a realistic electrochemical environment for studying Li dendrites nucleation.

In the classical MD framework, a straightforward way to achieve a ConstP simulation is to assign a charge to each atom on the electrode surface, with charge values being predetermined by the surface charge density at a specific potential and not updated along MD trajectories. This approach thus can introduce a constant electric field across two electrodes. For example, Wu et al.[15] applied this method to investigate the effect of the electric double layer on electrolyte decomposition and solid electrolyte interphase (SEI) formation. However, to effectively account for more complex electrochemical interfaces, the fixed potential method[16,17,18] is a more popular way to achieve the ConstP condition. This method adds an external potential to the electrode, and by further solving Poisson's equation or adopting the charge equilibration (QEq) method[19,20], atomic charges can be automatically solved given a specific configuration. For example, reactive force field (ReaxFF)[21,22,23] MD simulations combined with the EChemDID[24] method have been employed to unravel the dendrites growth dynamics in Li metal batteries.[25,26] Nevertheless, classical force field-based MD simulations are often constrained by their limited accuracy, particularly in complex electrochemical systems. In the *ab-initio* MD (AIMD) framework, the so-called grand canonical density functional theory (GC-DFT) method[27,28,29,30] can achieve a fixed work function (or potential equivalently) condition by iteratively varying total electron numbers of electrode surface systems, although not conforming to an exact GC distribution of microstates. Moreover, AIMD simulations also suffer from large computational cost, which limits their applications in complex battery interfaces. Balancing the trade-off between efficiency and accuracy, machine learning force field (MLFF)[31,32,33] has been found extensive applications in the LiB research.[34,35,36,37,38,39] Although several ConstP methods under the MLFF framework have been developed[40,41], they cannot provide variable atomic charges due to lack of physically meaningful solvation of Coulomb interaction, which limits their accuracy in modeling Li electrodeposition/dissolution processes. Therefore, it is highly desired to develop an efficient and accurate ConstP method under the MLFF to elucidate Li electrochemical behaviors during the charging/discharging cycles.

In this work, by combining the MLFF with the QEq method, we develop a computational scheme for performing high-efficiency ConstP MD simulations. We apply the proposed method in modeling the electrochemical interface of the ethylene carbonate + lithium hexafluorophosphate [EC+$LiPF_6$] electrolyte in contact with Li metal electrodes (denoted as Li/[EC+$LiPF_6$] throughout this paper), and directly visualize and investigate the dendrites nucleation during charging/discharging processes. Our simulations successfully present the dynamic process of dendrites nucleation, and show that inhomogeneous Li depositions following Li aggregations can initiate dendrites nucleation. Quantitative analysis of our MD trajectories further provides microscopic insights that the local aggregation of Li atoms in amorphous inorganic SEI components is the key mechanism triggering inhomogeneous Li depositions. Overall, our proposed scheme provides an efficient and accurate simulation tool for modeling realistic ConstP conditions in complex electrochemical interfaces in batteries.

## Results and Discussion

### Methodological principles

We first introduce the principles of implementing constant charge (ConstQ) and ConstP constraints under the MLFF framework in this work. More details will be shown in Supporting Information (SI). The architecture is schematically presented in **Figure 1a**, where the total potential energy $E_{\mathrm{Total}}$, derived from DFT calculations, is partitioned into two parts:

$$E_{\mathrm{Total}} = E_{\mathrm{Short}} + E_{\mathrm{QEq}} \quad (1)$$

where the complex atomic interaction consists of a long-range component and a short-range component, and the long-range part can be approximated by electrostatic interactions, for which we employ the QEq method to describe. We refer the residual part (i.e. $E_{\mathrm{Total}} - E_{\mathrm{QEq}}$) as the short-range component, which will be represented by MLFF in our work. The $E_{\mathrm{Short}}$ term thus mainly represents the local bonding interactions between atoms.[42] In fact, this idea of partitioning the total energy into a long-range component and a short-range component has been extensively adopted in previous MLFF models, such as the CENT[43], 4G-HDNNP[44], BAMBOO[45], SpookyNet[46], SO3LR[47] and so on. However, we note that all these models cannot be employed to handle realistic electrochemical reactions under a ConstP condition. In the QEq method, the atomic charges of modeling systems containing $N$ atoms are optimized to minimize the QEq energy[20]

$$E_{\mathrm{QEq}} = E_{\mathrm{Coulomb}} + \sum_{i=1}^{N} \left( \chi_i^0 Q_i + \frac{1}{2} J_i Q_i^2 \right) \tag{2}$$

where $E_{\mathrm{Coulomb}}$ represents the Coulomb interactions of Gaussian charges, which can be computed by considering point charge interactions using the Particle Mesh Ewald (PME)[48] algorithm (Eq. S2 in SI), and further adding a Gaussian charge distribution correction term (Eq. S6 in SI). $Q_i$, $\chi_i^0$, and $J_i$ are the atomic charge, electronegativity and hardness of each atom in the modeling system, where the subscript "$i$" refers to atomic index. Atomic charges $Q_i$ will be computed by $E_{\mathrm{QEq}}$ minimization for every single configuration, and parameters of $\chi_i^0$ and $J_i$ of each element species are set up (see Table S1) at the beginning of molecular simulations.

In the ConstQ constraint, the sum of all atomic charges is constrained to the total charge $Q_{\mathrm{tot}}$ of the system ($\sum_{i=1}^{N} Q_i = Q_{\mathrm{tot}}$), we thus employ the Lagrange multiplier method to solve this constrained minimization problem

$$\mathcal{L} = E_{\mathrm{QEq}} - \chi_{\mathrm{eq}} \left( \sum_{i=1}^{N} Q_i - Q_{\mathrm{tot}} \right) \tag{3}$$

The QEq charges $\{Q_i\}$ and the Lagrange multiplier $\chi_{\mathrm{eq}}$ can thus be solved by

$$\begin{cases} \dfrac{\partial \mathcal{L}}{\partial Q_i} = 0 \\ \dfrac{\partial \mathcal{L}}{\partial \chi_{\mathrm{eq}}} = 0 \end{cases} \tag{4}$$

In the ConstP condition, we apply external potentials to electrode atoms with a set of predefined values $\{\phi_i\}$, and the grand energy of this system is given by

$$\Omega = E_{\mathrm{QEq}} + \sum_{i=1}^{N} \phi_i Q_i \tag{5}$$

where

$$\phi_i = \begin{cases} 0, & \text{if not Li} \\ 0, & \text{if Li, and } \mathrm{CN}_{\mathrm{Li-Li}} < \mathrm{CN}_{\mathrm{Metal}} \\ \phi_{\mathrm{Li},1}, & \text{if Li, } \mathrm{CN}_{\mathrm{Li-Li}} > \mathrm{CN}_{\mathrm{Metal}}\text{, and Li} \in \text{Anode side} \\ \phi_{\mathrm{Li},2}, & \text{if Li, } \mathrm{CN}_{\mathrm{Li-Li}} > \mathrm{CN}_{\mathrm{Metal}}\text{, and Li } \in \text{Cathode side} \end{cases} \tag{6}$$

Here, $CN_{Li\text{-}Li}$ means the coordination number (CN) of a Li atom with surrounding Li atoms in simulations, and $CN_{Metal}$ represents the CN of an atom (e.g. Li) in the corresponding bulk metal phase. Considering the electrode material is the body-centered-cubic (BCC) Li metal in our work, $\mathrm{CN}_{\mathrm{Metal}}$ is set as 8. In principle, we simulate an open system under the ConstP condition, which means a net charge may persist during simulations. Because the whole cell must maintain neutrality required by the periodic boundary condition, we thus add the neutral constraint ($Q_{\mathrm{tot}} = 0$) in the ConstP condition. The Lagrangian should be modified as

$$\mathcal{L} = \Omega - \chi_{\mathrm{eq}} \left( \sum_{i=1}^{N} Q_i - Q_{\mathrm{tot}} \right) \tag{7}$$

The QEq charges $\{Q_i\}$ and the Lagrange multiplier $\chi_{\mathrm{eq}}$ can also be solved by using Eq. 4. From above derivations, we can see that introducing an external potential is mathematically equivalent to shifting the atomic electronegativity. A positive/negative $\phi_i$ value enhances the reducing/oxidizing property of the Li metal electrode in this work, which corresponds to a cathode/anode in our interfacial model. By assigning distinct external potential values to atoms belonging to different electrodes in an electrochemical cell, charges on electrode atoms can be updated along MD trajectories to simulate a pair of counter electrodes (including both a cathode and an anode) in a single modeling supercell.

The presence of atomic charges may induce a non-negligible dipole moment, especially along the direction perpendicular to the interface (denoted as the *z* direction in this work), the dipole correction term[49] should be added

$$E_{\text{corr}}^{\text{dipole}} = \frac{2\pi}{V}\left(M_z^2 - Q_{\text{tot}}\sum_{i=1}^{N} Q_i z_i^2 - Q_{\text{tot}}^2 \frac{L_z^2}{12}\right) \quad (8)$$

$$M_z = \sum_{i=1}^{N} Q_i z_i \quad (9)$$

where $L_z$ is the box length along the direction perpendicular to the electrode/electrolyte interface and $z_i$ denotes the *z* component of *i*-th atomic coordinates. This correction therefore should be included in the $E_{\text{QEq}}$ term. Since $Q_{\text{tot}} = 0$ as required by the neutrality of the periodic supercell, $E_{\text{corr}}^{\text{dipole}}$ here only depends on the total dipole moment of the simulation cell.

The idea of extracting the short-range interaction term $E_{\text{Short}}$ from DFT calculations based on QEq calculations enables us to employ a MLFF framework to represent $E_{\text{Short}}$. The corresponding QEq calculations need to be performed under a ConstQ condition, which is consistent with typical DFT setups. This also enables us to train exclusively on the short-range component, while implementing a ConstP condition through electronegativity shifts to metallic atoms, resulting in a significantly reduced training cost, compared to earlier studies[40,41] conducting separate trainings under multiple bias potentials (equivalent to fitting multiple potential energy surfaces). Once QEq charges (at ConstQ) are solved, we can obtain the QEq energy and forces. The short-range component $E_{\text{Short}}$ then can be computed by

$$E_{\text{Short}} = E_{\text{DFT}} - E_{\text{QEq}} \quad (10)$$

which will be further used in training MLFF. Since we employ the deep potential (DP) model[50] to describe the short-range component in this study (referred as $\text{DP}_{\text{Short}}$), we denote this new approach as the **DP-QEq** method. For the clarity of following discussions, the traditional DP model fitting the total DFT energies is termed as the "full DP".

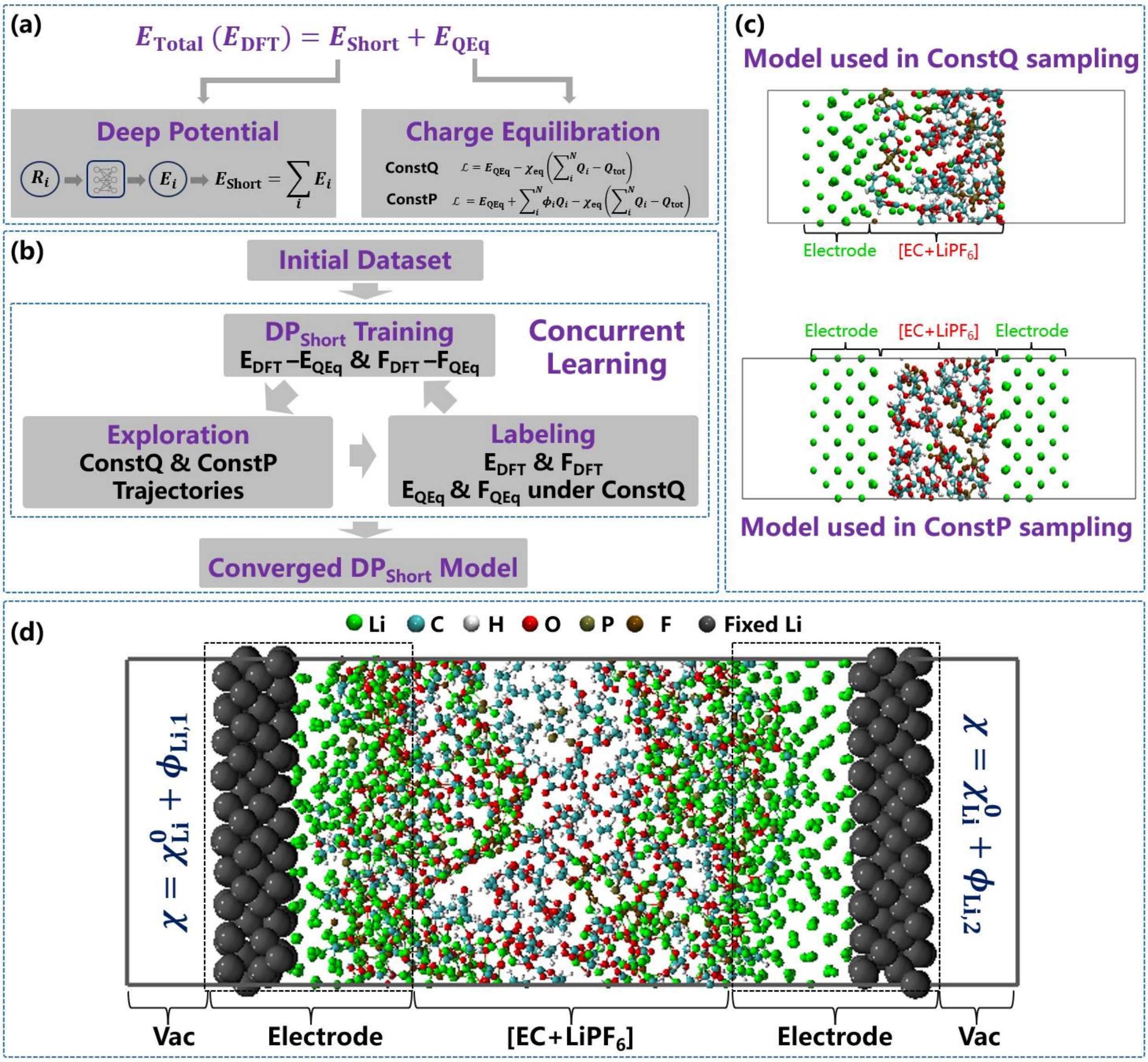

**Figure 1.** (a) The overall architecture of the DP-QEq model. (b) The workflow of automatic dataset generation along ConstQ and ConstP trajectories for training the $DP_{Short}$ component used in our DP-QEq approach. (c) Interface models used in the ConstQ and ConstP samplings for generating training dataset (details discussed in Method section). (d) Schematic atomic structure of the interface model for studying the dynamics of Li dendrites nucleation under realistic electrochemical conditions.

## Performance of the trained MLFF model

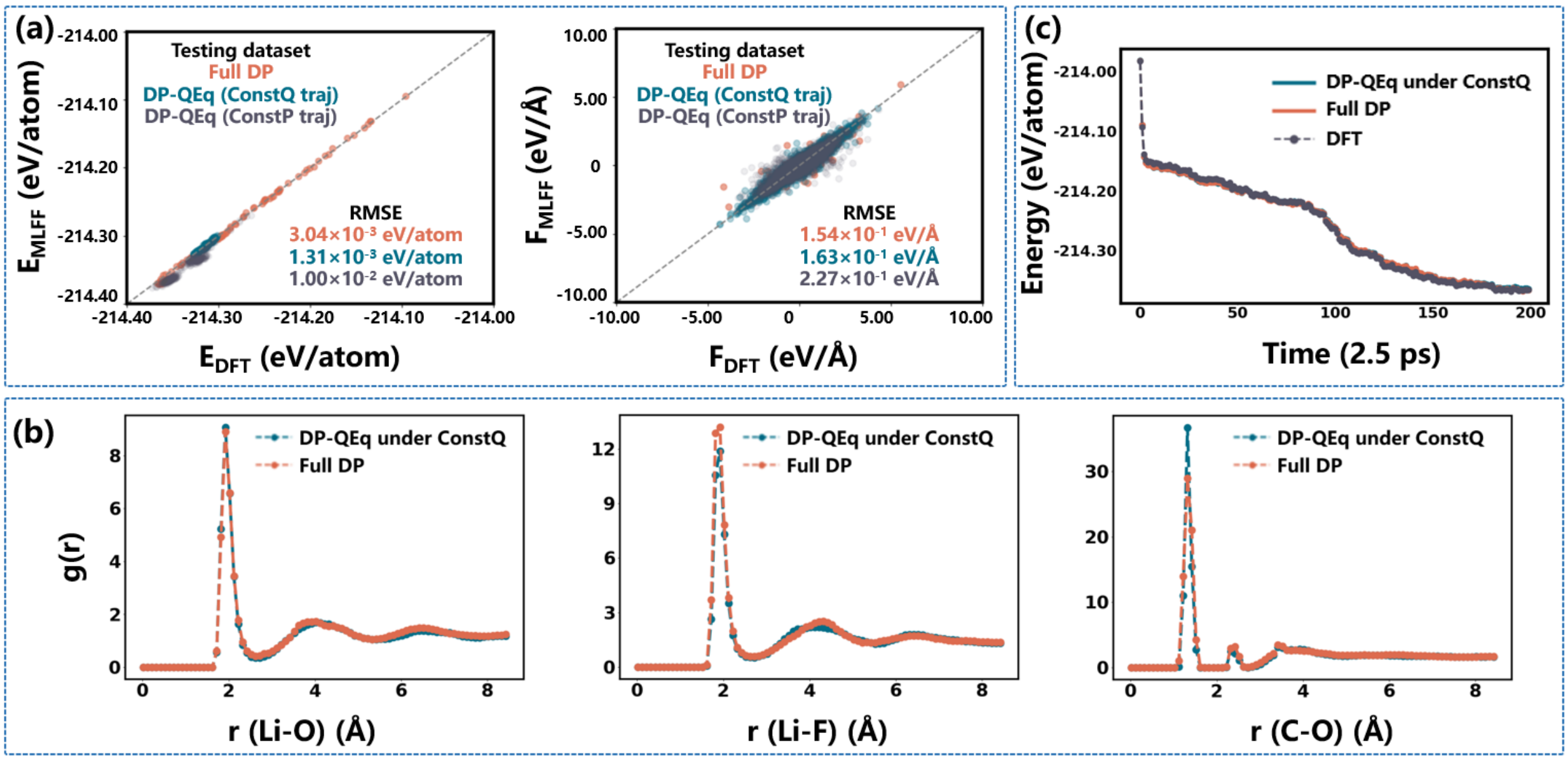

**Figure 2.** (a) Comparisons of the energies and forces obtained from DFT, full DP and DP-QEq models on the testing dataset, which consists of a 50 ps ConstQ trajectory and 4 rounds of 100 ps ConstP trajectories for an interfacial model shown in the lower panel of Figure 1c. The red, blue and gray numbers represent the RMSE values between full DP vs DFT, DP-QEq under ConstQ vs DFT, and DP-QEq under ConstP vs DFT, respectively. (b) Comparisons of the Li-O, Li-F and C-O pairs' RDF curves obtained from a 500 ps full DP MD trajectory and a 500 ps DP-QEq MD trajectory under a ConstQ condition. (c) Comparisons of energies along a DP-QEq driven 500 ps MD trajectory under a ConstQ condition. The DP-QEq, full DP and DFT methods are employed to compute energies of each configuration belonging to this trajectory. The cell shown in the lower panel in Figure 1c is used as the initial configuration for this 500 ps ConstQ MD simulation.

We first train a full DP model (training parameters are same as the $DP_{Short}$, see the Methods section) using the dataset calculated by the DFT method, which, as we discuss below, can help us validate the $DP_{Short}$ model under the ConstQ condition. We can see that the energies and forces predicted by the full DP model closely align with the DFT results both on the training dataset (Figure S1-S3) and testing dataset (details of the testing dataset generations can be found in section 2.2 of SI, Figure S4). The total root mean square errors (RMSEs) of energy and forces on the testing dataset are 3.0 meV/atom and 0.15 eV/Å, respectively, indicating the robustness of our trained full DP model (**Figure 2a**). We then train a $DP_{Short}$ model by subtracting the long-range QEq part from the DFT results (see Figure S5-S7). We further construct a testing dataset for the $DP_{Short}$ model, consisting of a 50 ps CosntQ trajectory and four rounds of 100 ps ConstP trajectories for an interfacial model shown in the lower panel in **Figure 1c**. The last snapshot of the 50 ps ConstQ simulation trajectory is used as the initial configuration for subsequent ConstP MD calculations. RMSEs of DP-QEq energies and forces (contribution from both $DP_{Short}$ inferences and QEq calculations) on its ConstQ trajectory, compared with the accurate results of $E_{DFT}$ and $F_{DFT}$, are 1.3 meV/atom and 0.16 eV/Å, respectively. More accurate energy predictions of the $DP_{Short}$ model on its ConstQ trajectory, compared to the full DP model (also on its own ConstQ testing trajectory), validate the reliability of extracting long-range components from the total energy. Although RMSEs of DP-QEq energies and forces on its ConstP trajectory exhibit an increase of errors, considering more variable configurations being involved in its 400 ps long-time testing trajectories, these values are within an acceptable error range by comparing with earlier battery studies using machine-learning potentials.[51,52,53,54] To further specify the force error origin, especially to check out the atomic force errors on Li metal slab and the nucleated Li dendrite which are the most relevant regions to our scientific topic, we first identify metallic Li atoms (including the Li electrode and the emerged Li dendrite atoms) within the SEI along the 4$^{th}$ round of ConstP trajectory and track the force errors of those Li atoms, in terms of RMSE compared to DFT-predicted forces. As shown in Figure S8, the force RMSEs of these Li atoms involved in Li dendrite nucleation are quite small (around 0.07 – 0.09 eV/Å) throughout the entire testing trajectory, indicating results of the dynamic process of Li dendrite formation observed in our work (see below) are physically meaningful. In addition, we perform two 500 ps MD simulations by the full DP and DP-QEq models under the ConstQ condition, with a pair of electrodes included in the cell (the lower panel shown in **Figure 1c**), and compute radial distribution function (RDF) comparisons for the final 100 ps trajectories (**Figure 2b**). We can see that the Li-O, Li-F and C-O pairs' RDF profiles generated by the DP-QEq method well reproduce the results of the full DP simulation, justifying the performance of our DP-QEq scheme in structure prediction. Furthermore, we relabel the potential energies for 200 configurations, sampled at 2.5 ps interval along the DP-QEq trajectory, using both the full DP and DFT methods. As shown in **Figure 2c**, the potential energies obtained by the DFT method and by the full DP model are in good agreement with those inferred from the DP-QEq force field. We also estimate the overall force RMSEs (full DP vs DFT and DP-QEq vs DFT) as a function of timestep along this 500 ps ConstQ trajectory (see Figure S9a) and we can see that the RMSEs in both the full DP and DP-QEq models remain consistent with the results shown in **Figure 2a**. We further decompose the force errors into every atomic type (see Figure S9b and S9c), and find that the maximum force errors appear on the P atoms. The relatively small force errors on Li atoms, which are most relevant to our investigated topic "Li dendrite formation", further indicate that our simulated results are meaningful. It is noteworthy that the MD simulation time in this test (500 ps) significantly exceeds that of the configurational exploration (20 ps) during the concurrent learning iterations in the $DP_{Short}$ training process, indicating a stable and robust performance of the $DP_{Short}$ model on MD testing trajectories.

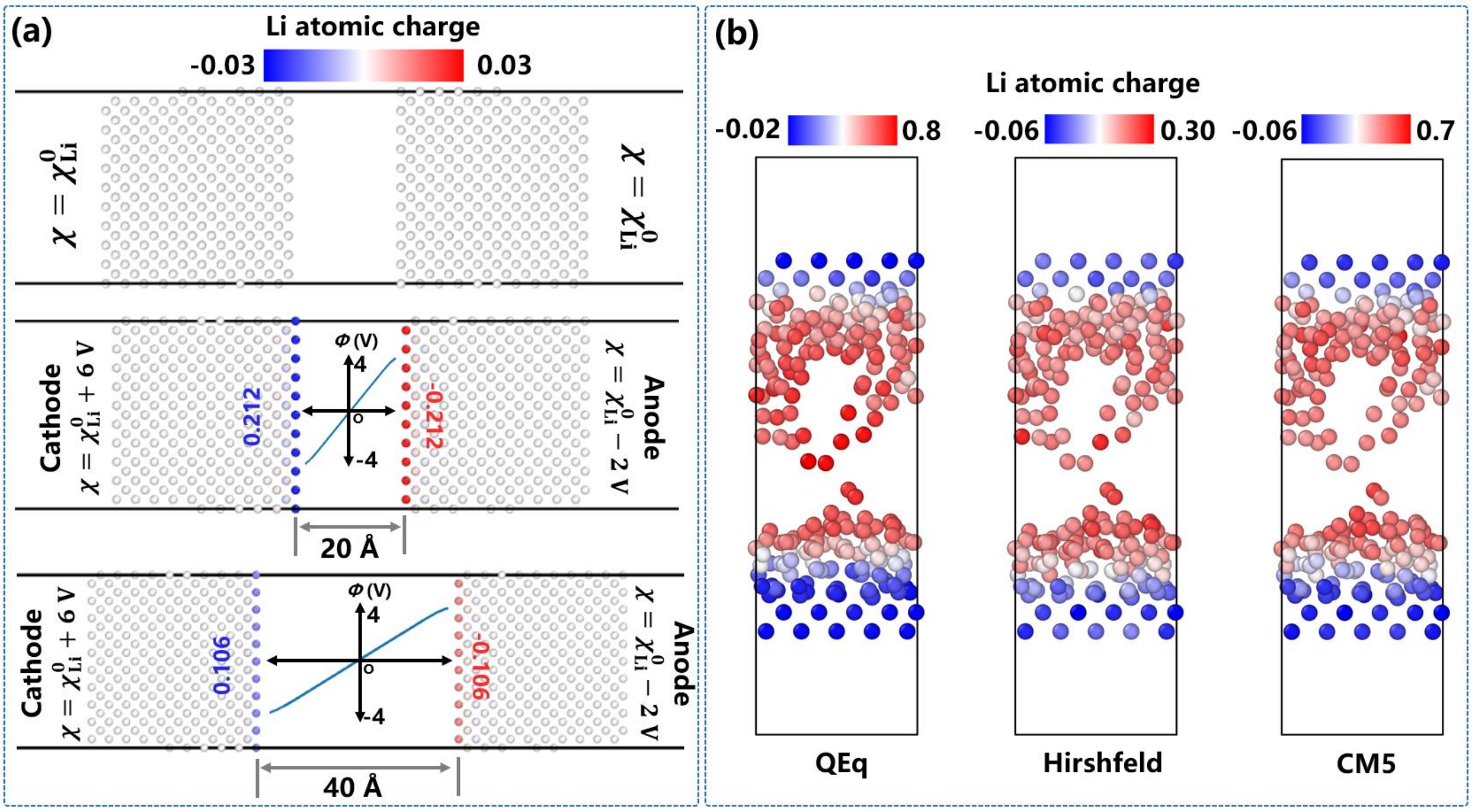

**Figure 3.** (a) Validations of the QEq method on a toy model, which consists of two Li metal electrodes separated by a vacuum. Schematic models from top to bottom represent: a ConstQ model with a distance of 20 Å between electrode slabs, a ConstP model with a distance of 20

Å between slabs with different bias potentials (electronegativity shifts) applied on two electrodes, and a ConstP model with a distance of 40 Å between slabs (bias potential setup consistent with the second model). Li atoms are color-coded by their atomic charges. The electrode surface charge densities (units in e/nm$^2$) are also illustrated in the plot (bule and red numbers). The electrostatic potential profiles are also plotted in the vacuum region (the potential value at the mid-point referred as 0 V). (b) Comparison of Li atomic charges predicted by the QEq method with those from DFT calculations for the Li/[EC+$LiPF_6$] interfacial model, where Hirshfeld[55] and CM5[56] post-processing charge analysis schemes are employed to output the Li atomic charges from DFT results. We only display the Li atoms for clarity, which are color-coded by their atomic charges.

## Validations of the DP-QEq method under ConstQ and ConstP conditions

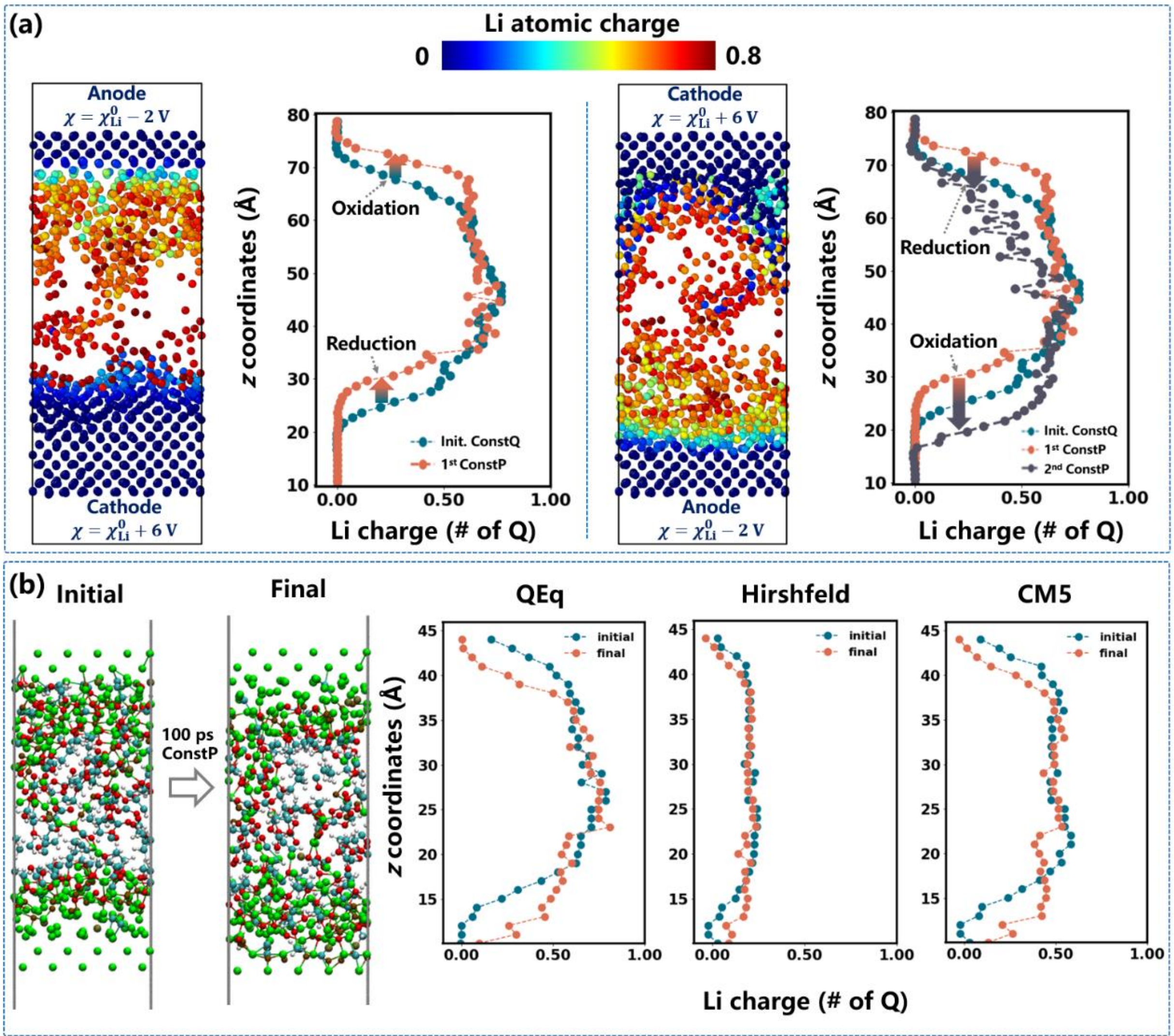

**Figure 4.** (a) Morphologies obtained from two rounds of 300 ps ConstP MD simulations. We only display the Li atoms for clarity, which are color-coded by their atomic charges in structural plots. The Li atomic charge distribution curves as a function of *z* coordinates are shown by the side. A positive/near-zero value means a Li ion/atom, "# of Q" means the quantity of unit charge Q. Init. ConstQ denotes the initial 500 ps ConstQ simulation. (b) The initial and final morphologies obtained from a 100 ps DP-QEq ConstP simulation for a small cell displayed in the lower panel in **Figure 1c**. The Li atomic charge distribution curves are shown by side for these two configurations, calculated by the QEq (a snapshot extracted from the ConstP trajectory, and charges calculated under the ConstQ condition) and DFT (charge analysis done by the Hirshfeld and CM5 methods), as a function of *z* coordinates.

We first validate the QEq method by a toy model, which consists of two Li electrode slabs separated by a vacuum (see **Figure 3a**). Given that all atomic electronegativities of two electrodes are identical under the ConstQ condition, the calculated atomic charges are zero, both on the surface or in the bulk (the top panel in **Figure 3a**). This is expected because there is no potential drop across the electrodes. However, the introduction of bias potentials (+6 V electronegativity shift for the cathode and -2 V electronegativity shift for the anode) leads to appearance of opposite net charges on the two electrode surfaces (the whole cell still neutral), while the charge within the electrode bulk interior region remains zero. This result makes physical sense because net charge cannot exist in the interior region of metals. The calculated surface charge densities are ±0.212 e/nm$^2$ (the middle panel in **Figure 3a**), closely matching the setup in a previous work[15] about the effect of electric double layer on SEI formation in Li batteries. We then increase the electrode's distance from 20 Å to 40 Å and observe that the electrode surface charge densities decrease exactly by a half (±0.106 e/nm$^2$, the bottom panel in **Figure 3a**). As expected, doubling the distance between the electrodes will result in halving the electric field strength to maintain a constant potential drop. We further investigate the electrostatic potential profiles between two electrode slabs by the probing charge method (more calculation details shown in SI Section 1.4) and find that the electrostatic potential difference between two electrodes in the toy model exactly equals to the difference of electronegativity shifts applied on Li atoms belonging to different electrodes. More importantly, the shape of those profiles indicates there is a uniform electric field in the middle region between two electrodes (the middle and bottom panels in **Figure 3a**), which match with analytical results of a typically parallel plate capacitor. All above validation simulations show a reasonable prediction of charge distributions in an ideal toy model under both ConstQ and ConstP conditions, and further demonstrate that the QEq method is capable of achieving a constant potential drop across the pair of electrodes.

Another critical aspect that needs to be verified is the qualitative validity of atomic charge values derived from the QEq method. To assess this, we compare the QEq charges with those from the *ab-initio* DFT method. Here we note that one possible arbitrariness of the so-called *ab-initio*-derived charges depends on different post-processing charge analysis schemes, we consider the Hirshfeld[55], CM5[56], and Bader[57,58] analysis methods in this section, using a small size Li/[EC+$LiPF_6$] interface model as depicted in the bottom panel of **Figure 1c**. Bader charges are obtained by using Bader[57,58] program, and Hirshfeld and CM5 charges are calculated by Multiwfn[59,60] program. We find that the discrepancies among different charge analysis schemes are quite significant, consistent with the observations reported by Choudhuri et al.[61] Particularly, the Bader charge model yields unphysical predictions, as unexpected non-zero net charges emerge within the interior region of metallic electrodes. In contrast, both the Hirshfeld and CM5 charge analysis schemes provide qualitatively reasonable charge results from DFT calculations, based on the observation that the charges within the Li metal interior region are predicted to be ~ 0 by Hirshfeld and CM5 (**Figure 3b** and Figure S10a). We thus present the comparison of charge prediction between the QEq method and the DFT-based Hirshfeld and CM5 analysis in **Figure 3b.** Details of the Bader charge results can be referred to SI Section 2.6. We can see from **Figure 3b** that the charge distribution generated by the QEq method qualitatively match well with the *ab-initio* charge results generated by Hirshfeld and CM5 post-processing analysis, justifying the validity of the QEq approach in terms of electrostatic interaction modeling.

We further validate our DP-QEq approach by a realistic model, which consists of two Li metal electrode slabs separated by electrolyte. We initially conduct a 200 ps ConstQ NPT MD simulation with the cell size being able to relax to avoid cavity generations caused by interfacial side reactions between the Li metal and electrolyte molecules, and then perform a 500 ps ConstQ NVT MD to generate a configuration used for subsequent ConstP simulations. Due to the high reactivity of Li metals, Li atoms on both surfaces tend to be oxidized (red color in the left panel of Figure S11a, indicating a cation state of Li) and react with the electrolyte to form an initial SEI layer, which can be confirmed by monitoring the molecular number of various chemical species over time steps. We identify chemical species using the ReacNetGenerator[62] code in this work. As shown in Figure S11b, the decreasing molecular numbers of EC and $PF_6$ indicate their decompositions, accompanied by formation of inorganic products ($Li_2CO_3$, $Li_2O$ and LiF) and gas molecules ($C_2H_4$, $CO_2$ and CO). The formation mechanisms of various products are shown in Figure S12. These products also have been found in earlier AIMD simulations[63] and experimental studies[64,65], further validating the reliability of the DP-QEq model. Here, we note that all those decomposition products have been explicitly included in our training dataset and testing dataset (see Table S3), and our DP-QEq model can thus accurately capture those chemical products (Figure S13). We can see that the inorganic components are accumulated and non-uniformly distributed at the bottom of the SEI layer (closer to the electrode surface, Figure S11b). Those findings are consistent with previous experimental and computational observations.[7,66,67,68,69] As a result, atomic charges of Li atoms progressively increase from electrode regions to the electrolyte. Since we include a pair of counter electrodes in our model without introducing bias potentials in this ConstQ simulation, equivalent oxidation side reactions occur on both electrode surfaces, as evidenced by an almost symmetric charge distribution curve relative to center of the electrolyte region (Figure S11a).

We then verify the implementation of the ConstP condition in our DP-QEq approach.

As mentioned earlier, a reliable ConstP simulation method must be capable of describing Li electrochemical depositions/dissolutions and identifying different types of electrochemical reactions occurring at surfaces of a double-interface model (i.e., oxidation reactions on the anode and reduction reactions on the cathode). We perform a 300 ps ConstP DP-QEq MD simulation based on the previous 500 ps ConstQ NVT MD-derived final morphology, with the upper/bottom electrode as the anode/cathode (illustrated by the left panel in **Figure 4a**, the 1$^{st}$ ConstP round). Introducing an oxidizing bias potential lowers the electronegativity of the Li atoms in the anode side, thus driving their further oxidation. As expected, the remaining metallic Li atoms (atomic charge ~ 0) near the anode/SEI interface in the previous ConstQ simulation (denoted by blue color at the anode surface, Figure S11a) are further oxidized to Li ions and lose their BCC local structures. While for the cathode side, some previously oxidized Li ions (during the ConstQ run) adjacent to the cathode surface, which are denoted by red and green colors in Figure S11a, become reduced and redeposit on the electrode surface. These observations can be confirmed by an overall upward shift of the Li charge distribution curve (illustrated by the left panel in **Figure 4a**). We subsequently reverse the electrode (the upper electrode as the cathode and the bottom electrode as the anode) and perform another 300 ps ConstP MD simulation (the final state of the previous ConstP trajectory as the initial state, the right panel in **Figure 4a**, the 2$^{nd}$ ConstP round). The electrochemical redox reactions of Li ions/atoms near the cathode and anode surfaces exhibit similar behavior as discussed in the 1$^{st}$ ConstP trajectory. Along this 2$^{nd}$ ConstP simulation, the Li charge distribution curve shifts downward as expected. To verify the above shifting behavior of Li charge distribution curves, we also conduct a round of 100 ps ConstP MD simulation on a small size cell (shown in the lower panel of **Figure 1c**), and use DFT calculations combined with the Hirshfeld and CM5 post-processing method to compute Li atomic charges of the initial and finial snapshots (**Figure 4b**). The charge distribution curves generated by both of Hirshfeld and CM5 schemes exhibit similar shifting behaviors from the initial to the final states, consistent with the predictions of our DP-QEq method (**Figure 4b**). Here, we emphasize that the reduction reaction occurring on the cathode side is $Li^+ \rightarrow Li^0$, which is different from the toy model with emergence of $Li^-$. The key difference between the ideal vacuum model and the realistic model with electrolyte in between lies in the absence vs. presence of $Li^+$ in the electrolyte environment near electrode surfaces. More discussions can be found in SI section 2.10. Furthermore, we also investigate the impact of the magnitudes of different bias potentials on the electrochemical Li redox behavior. As shown in Figure S14, a more negative or more positive bias potential will accelerate the Li oxidation or reduction reactions.

Overall, all above validations demonstrate that our proposed DP-QEq method under a ConstP condition enables cyclic simulations of Li electrochemical redox reactions in an interfacial supercell involving both cathode and anode surfaces.

## Atomic insights into Li dendrites nucleation dynamics

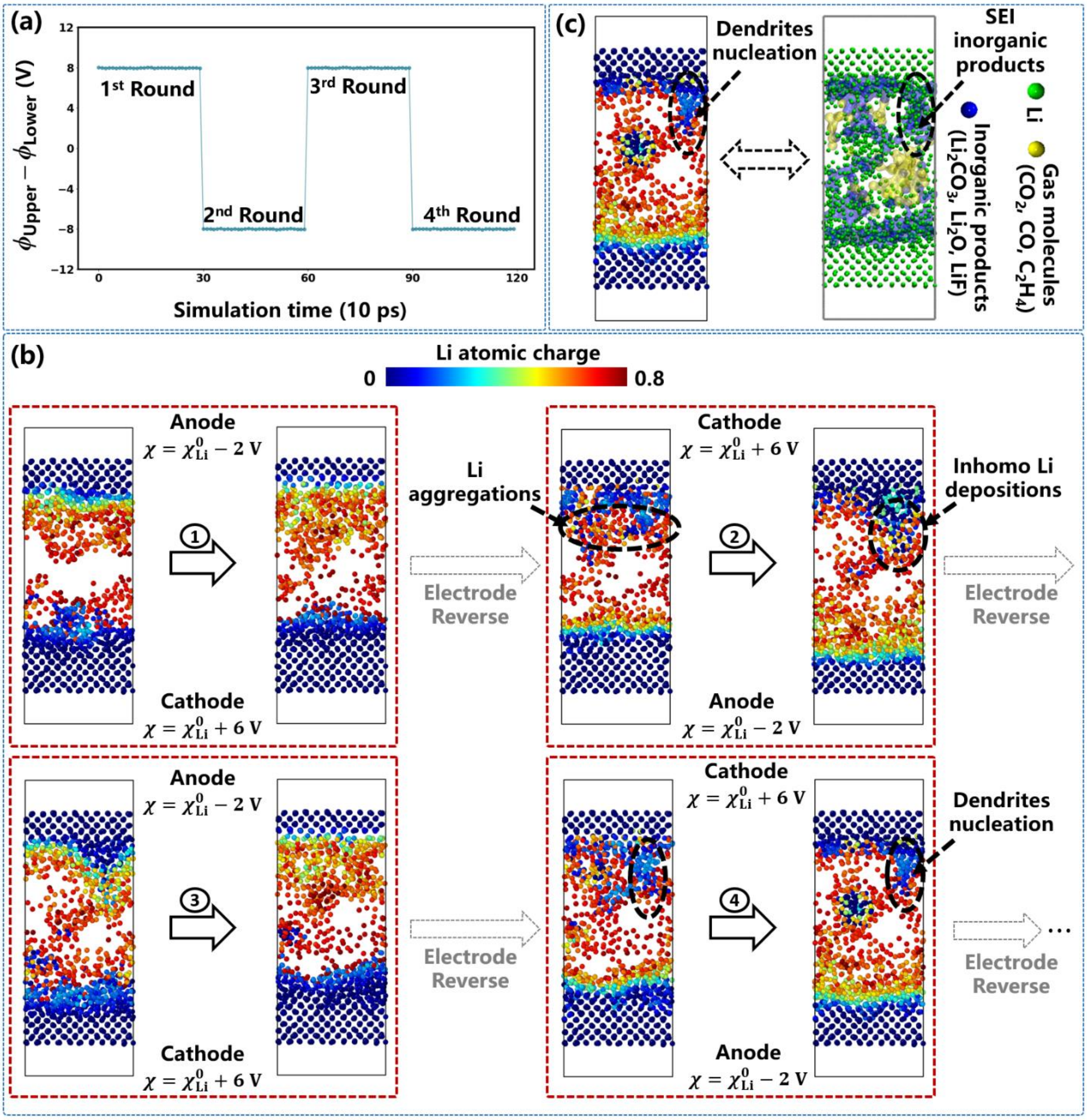

**Figure 5.** (a) Evolutions of the electrostatic potential (computational details shown in SI section 1.4) difference between the upper and lower electrodes with respect to the simulation time. (b) Charge state distributions of Li atoms/ions during the whole cyclic ConstP MD simulations. Our interface model involves a pair of counter electrodes where $Li^+ \rightarrow Li^0$ reduction reactions occur at the cathode surface, and vice versa. Only the Li atoms are shown in the plots for clarity. We can switch the anode and cathode sides by reversing the applied potentials on the upper and bottom Li electrodes. Each red dashed box represents one round of 300 ps ConstP MD simulation with a solid arrow inside pointing from the initial state to the final state along the corresponding trajectory. The circled number on top of a solid arrow denotes the round's serial index. Dashed arrows indicate the reverse of potentials applied on electrodes. The path of inhomogeneous Li depositions at electrode surface is highlighted by black dashed circles. (c)

Distributions of (left) the Li atomic charge state and (right) various products in SEI for the final snapshot of the 4th round of ConstP MD simulation shown in Figure 5b. We only display the Li atoms (green), inorganic products (blue), and gas molecules (yellow) for clarity in the right panel. Visualizations are made by the VMD software[70]. The inhomogeneous Li deposition regions are highlighted by black dashed circles.

To explore dendrites nucleation dynamics, we perform four rounds of ConstP MD simulations with a 500 ps ConstQ MD-derived morphology as the initial structure, during which we reverse the applied potential values on the pair of Li electrodes successively to mimic cyclic Li electro-depositions/dissolutions at cathode and anode surfaces along electrochemical cycles of Li batteries. Each ConstP MD round lasts for 300 ps (i.e., $3\times10^5$ MD steps). **Figure 5a** shows the potential difference (between the two electrodes) fluctuation with respect to MD steps along whole trajectories. We find that the electrostatic potential difference between two electrodes in the ConstP simulations exactly equals to the difference of electronegativity shifts applied on Li atoms belonging to different electrodes, demonstrating the good performance of our DP-QEq model in describing electrostatic properties for achieving an electrochemical condition. **Figure 5b** illustrates the charge distribution evolutions of Li atoms during sequential electrochemical runs. The Li charge distribution information enables us to monitor the Li atoms' redox states and analyze their electrochemical deposition/dissolution behaviors. Our simulations reveal that inhomogeneous Li depositions, following Li aggregations in SEI, can trigger the dendrites nucleation. Specifically, Li atoms at the anode surface are oxidized and react with the EC or $LiPF_6$ to form a SEI layer in the 1st round simulation. Upon reversing the potentials applied on the electrodes, we observe the emergence of Li aggregations (denoted by blue balls) within the SEI layer (Figure S15 and the left panel in the 2nd round in **Figure 5b**). These non-uniform distributed Li aggregations near the cathode surface could induce inhomogeneous depositions, initiating Li dendrites nucleation (the right panel in the 2nd round in **Figure 5b**). Li dendrites continue to grow and become relatively sharp in subsequent simulations (the right panel in 4th round in **Figure 5b**). Meanwhile, the tip of the dendrite strips to form an isolated Li cluster, which is ~ 1 nm away from the dendrite (the right panel in $4^{th}$ round in **Figure 5b**). We realize that the cutoff radius for counting $CN_{Li\text{-}Li}$ is an important parameter that determines the dimension of a local space where a Li atom could feel a metallic environment. Li dendrite formation along our DP-QEq ConstP MD simulations might depend on the cutoff radius setup, we thus need to test this parameter to ensure the reproducibility of our simulation findings. A similar Li dendrites nucleation process is observed with a 3.3 Å $CN_{Li\text{-}Li}$ cutoff (see Figure S16), justifying the setup parameter in our calculations.

At the end, we unravel the origin of dendrites nucleation by visualizing distribution of various chemical species (mainly including inorganic products and gas molecules) in SEI. We plot the Li atomic charge distribution of the final snapshot in the $4^{th}$ round of ConstP DP-QEq simulation and the corresponding products' distributions of this snapshot in **Figure 5c** to clarify the spatial correlation between the nucleation regions and SEI products. We can see that the SEI exhibits a significant heterogeneity, with inorganic products ($Li_2CO_3$, $Li_2O$ and LiF) locating at bottom of the SEI region (closer to electrodes), while gas molecules ($CO_2$, CO and $C_2H_4$) are dispersed more randomly. This nonuniform characteristic of the interfaces matches well with experimental observations.[64,69,71,72,73] Here, we note that dendrites nucleation is less likely to be affected by SEI organic products' distribution[7], which thus is not discussed in this study. We first find that Li aggregation appears in dendrites nucleation region, which can be evidenced by the increase of Li-Li bond numbers in this region in the last round compared to the initial round (see Figure S17). The Li-Li distances for the Li ions dissolved in electrolytes are relatively large due to the presence of solvation structures, making it difficult to form Li-Li bonds effectively. However, once the electrolyte and Li salts react with Li metals forming SEI layers, the Li-Li distance is significantly reduced in the inorganic components (e.g. LiF and $Li_2O$), especially for the amorphous morphologies, which could lead to local Li ion oversaturation[74] and nonuniform distribution. We then find that Li inhomogeneous depositions at the cathode surface, which is clearly represented by the dark blue region in the left panel of **Figure 5c**, directly relates to Li dendrite nucleation. More importantly, the region with Li dendrite spatially overlaps with a region of SEI inorganic products, which is highlighted by the dashed black circles in both panels of **Figure 5c**. Therefore, inhomogeneous Li depositions, following Li aggregations in amorphous inorganic components of solid electrolyte interphases, can initiate dendrites nucleation. Avoiding local Li-rich spots in SEI inorganic components therefore may mitigate inhomogeneous deposition from a material design's perspective. In fact, a recent experimental study demonstrated that elimination of nonuniform Li oversaturation by modulating inorganic products in SEI results in more homogeneous Li deposition,[74] which is consistent with our mechanistic insight provided in the above discussions.

## Conclusion and Outlook

In summary, we propose a ConstP approach under the MLFF framework that enables direct observation of Li electro-depositions/dissolutions during cyclic simulations. Utilizing this method, we successfully present the dynamic process of the Li dendrites nucleation during the electrochemical cycles, taking the Li/[EC+$LiPF_6$]/Li as the modeling double-interface system. Our simulations show that the predominant pathway of Li dendrites nucleation stems from the nonuniform Li deposition initiated by Li aggregations in the SEI inorganic components. Overall, our simulations reveal the microscopic dynamic processes in the early stage of Li dendrites nucleation. Moreover, we present an efficient and accurate simulation tool for modeling realistic electrochemical condition, achieving ConstP conditions under the MLFF framework. We thus expect our proposed scheme to be widely applied in complex electrochemical interfaces, such as structural and chemical evolvements of solid-solid interfaces in all-solid-state batteries along electrochemical operation cycles, or electrodes' surface corrosion in electrocatalytic systems, and so on.

Compared to earlier approaches for electrochemical reaction simulations, our DP-QEq

ConstP method exhibits several advantages. To the best of our knowledge, this is the first-time implementation of ConstP MD method with variable atomic charges within a MLFF framework, balancing both of computational efficiency and accuracy. In addition, we utilize the project gradient approach[75] combined with the automatic differentiation technique to solve the QEq charges (see SI Section 1.3). This combination exhibits a quasi-linear scaling and excels in handing large systems, thus enables simulations of a system containing thousands of atoms with long-range electrostatic interactions being well treated. This efficient computational solution has been integrated into the DMFF software[76]. Last but not the least, our DP-QEq method can precisely control the potential difference between two electrodes, which enables us to construct a more reasonable interface model, a double-interface model including a pair of counter electrodes, where the redox reactions occur simultaneously on both electrode surfaces. The existence of both an anode and a cathode in a single simulation cell makes it possible to balance the charge transfer just via the Li redox, i.e. the amount of $Li^0 \rightarrow Li^+ + e^-$ charge transfer at the anode side could equalize the amount of $Li^+ + e^- \rightarrow Li^0$ charge transfer at the cathode side, without enforcing the electrolyte to participate in redox side reactions (as conducted in several earlier studies[25,26] where only a single electrode contacts with electrolyte). Our double-electrode model thus provides a more realistic description for electrochemical cells.

We also need to point out that our proposed method still has several aspects that need to be further optimized. For example, atomic charges obtained from the QEq method may quantitatively differ from those generated by DFT calculations, as we discussed earlier. The parameters used in the QEq method, such as the electronegativity and hardness, can be inversely optimized to reproduce the DFT reference charges. In addition, the applied potentials, or equivalently the electronegativity shifts in our current model, can just provide redox driving force near cathode/anode surfaces, but cannot accurately match the actual operating voltage values of batteries, which is a direction for the future development of our method optimizations.

# Methods

## Setup details of DFT calculations

We performed all DFT calculations using the Atomic-orbital Based Ab-initio Computation at UStc (ABACUS) software (version 3.4.0).[77,78] We employed the Perdew-Burke-Ernzerhof (PBE)[79,80] functional and the linear combination of numerical atomic orbital (NAO) basis sets at the level of double-zeta plus polarization (DZP).[81] Specifically, we used $4s^1p$, $2s^2p^1d$, $2s^1p$, $2s^2p^1d$, $2s^2p^1d$ and $2s^2p^1d$ NAO for Li, C, H, O, P and F elements, respectively. We set the radius cutoffs for NAO of Li, C, H, O, P, and F as 7, 7, 6, 7, 7, and 7 Bohr, respectively. We employed the multi-projector "SG15-ONCV"-type norm-conserving pseudopotentials to describe the interactions between the nuclei and valence electrons. The valence electron configurations were $1s^2 2s^1$ for Li, $2s^2 2p^2$ for C, $1s^1$ for H, $2s^2 2p^4$ for O, $3s^2 3p^3$ for P and $2s^2 2p^5$ for F. We set the energy cutoff as 100 Ry. We included Grimme's D3BJ dispersion correction to account for vdW interactions.[82] We set the convergence criterions of the self-consistent-field calculation and structural optimizations as $10^{-6}$ Ry and 0.01 eV/Å, respectively. We used a 10×10×10 k-point mesh for the bulk body-centered-cubic (BCC) Li and a GAMMA-centered k-point mesh for the bulk [EC+$LiPF_6$] and Li/[EC+$LiPF_6$] interfacial models with relatively large supercell sizes.

## ConstQ and ConstP MD simulations

We performed all ConstQ and ConstP MD simulations by the Atomic Simulation Environment (ASE) package.[83] We defined a new calculator to update forces used for MD simulations. This new calculator returns the interaction terms contributed from both the short-range $DP_{Short}$ and the long-range QEq components (under either a ConstQ or a ConstP condition). We presented the workflow of MD simulations in Figure S18. Specifically, we first conducted a 200 ps NPT ConstQ MD simulation (pressures set as 1 bar along the *x* and *y* directions and as 100 bar along the *z* direction) for a superlattice model (3070 atoms, Figure S18a → S18b) to avoid cavity generations (induced by side reactions between Li metal and electrolytes) along MD trajectories. Given a sufficient thickness of the Li metal electrode, Li atoms in the slab's interior region remains a BCC structure after the above simulations. We then constructed a double-interface model containing a pair of counter electrodes as displayed in **Figure 1d** and Figure S18c (the box dimensions as 33.00×16.07×90.00 $Å^3$), by cutting from the middle of the Li metal slab in the above superlattice model. We further performed a 500 ps NVT ConstQ MD simulation to obtain an interfacial configuration for subsequent ConstP simulations. We then conducted four rounds of 300 ps DP-QEq NVT ConstP simulations by applying different bias potentials on the pair of Li metal electrodes, along which we successively reversed the potentials to mimic Li electrochemical redox cycles at cathode and anode surfaces. The final configuration of a previous round was employed as the initial structure setup for the next round. Here we need to note that one "round" represents a single ConstP simulation trajectory, and two rounds denote a cycle of ConstP simulations, between which the bias potentials on the pair of Li metal slabs are reversed once. Since we focus on capturing dynamic evolutions of the Li metal/electrolyte interfaces along successive electrochemical cycles, which are inherently non-equilibrium processes, it is unnecessary to achieve fully equilibrium states along each ConstP round, considering both the feasibility and rationality of our simulations. A trajectory of 300 ps for each ConstP round is already sufficient[25,26] to provide us dynamic insights of Li dendrites nucleation during these non-equilibrium processes at electrode/electrolyte interfaces. The uppermost/bottommost four layers of the top/bottom Li slab were fixed along those ConstP simulations to maintain a bulk environment for the interior regions of electrodes (Figure S18d). We used a timestep of 1 fs to integrate Newton equations of motion. We used the Nosé-Hoover[84,85] thermostat in all NVT simulations. As for the NPT simulations for relaxing cell volumes, we employed the Nosé-Hoover thermostat coupled with the Parrinello-Rahman dynamics[86] controlling the barostat. A 3-D periodic boundary condition was employed throughout all simulations.

We decided whether to apply a bias potential on a certain Li atom in our simulation cell based on its $CN_{Li\text{-}Li}$ with a cutoff of 3.5 Å during the ConstP MD runs (see Eq. 6). The electrode material is BCC Li metal in our work, $CN_{Metal}$ thus equals 8. If the $CN_{Li\text{-}Li}$ of a Li atom was larger than 8, an electronegativity shift would be applied to the corresponding Li atoms. Our previous validations on the toy model show that an electrode electronegativity difference of 8 V can yield a relatively reasonable surface charge density (matching the corresponding density in an earlier theoretical work[15]). Moreover, this relatively large bias potential difference between the counter electrodes could accelerate the Li redox reactions, enabling a practical simulation for observing Li dendrite formation within limited MD time scale. We therefore chose the electronegativity difference of 8 V (also similar with setups in earlier works[15,25,26]). For the specific values of $\phi_{\mathrm{Li},1}$ and $\phi_{\mathrm{Li},2}$, we found that the setup of $\phi_{\mathrm{Li},2} = +6$ V for the cathode side is an appropriate option for observing Li electrodepositions (more discussion can be found in SI Eq. S17), we thus set $\phi_{\mathrm{Li},1} = -2$ V for the anode side and $\phi_{\mathrm{Li},2} = +6$ V for the cathode side. The cathode and anode sides, when we consider applying opposite bias potential values on Li atoms within a single simulation supercell, were differentiated by the mid-line of the electrolyte region. We once again stress that although the applied potential values (represented by the electronegativity shifts as shown above) may deviate from the actual operating voltages of Li-ion batteries, it allows us to investigate the microscopic dynamics of Li dendrite nucleation within an affordable temporal scale by accelerating electrochemical reactions via an enhanced redox driving force.

## Dataset generations

**Initial dataset generations for training $DP_{short}$.** The initial training dataset consists of three parts: bulk Li, bulk EC and Li/[EC+$LiPF_6$] interface. For the bulk Li, we first obtained the structure of the BCC Li metal from the Materials Project database.[87] We then applied random perturbations on the atomic positions of the optimized structure and performed AIMD simulations to generate the initial dataset. For the bulk [EC+$LiPF_6$], we generated the initial random morphology by the Packmol[88] software, where the density of the EC is 1.3 g/cm$^3$ and the concentration of the $LiPF_6$ is 2.5 M. We performed a 200 ps classical ReaxFF simulation under 300 K to facilitate configurational sampling. We extracted structures along the trajectory at 100 fs interval for subsequent DFT calculations. For the Li/[EC+$LiPF_6$] interface system, we chose the Li metal (100) facet to construct the slab model, which contains 6 layers in the *z*-direction and the dimensions in the *x-y* plane were 17.15 Å×17.15 Å. For the [EC+$LiPF_6$], we also employed the Packmol[88] software to generate the random mixing configuration, where the dimensions in the *x-y* plane were fixed at 17.15 Å×17.15 Å to directly contact with the Li metal slab for the interfacial model construction (534 atoms, the upper panel in **Figure 1c**). We then performed a 200 ps ReaxFF MD simulation and extracted structures at 100 fs interval for subsequent DFT calculations. The ReaxFF simulations were carried out using the LAMMPS[89] software. We note here that these ReaxFF simulations were employed just to construct a reasonable initial dataset that covers a larger configurational space, which can help accelerate the convergence of the subsequent concurrent learning (configurational exploration not driven by ReaxFF anymore). After finishing DFT labeling calculations, the QEq energies and forces (under a ConstQ constraint) will be subtracted from DFT results to generate the initial $DP_{Short}$ force field model.

**Concurrent learning iterations.** We performed the concurrent learning processes by the DPGEN[90] workflow, which contains a series of iterations (see **Figure 1b**). Each iteration is composed of three steps: training, exploration and labeling. In the training stage, four $DP_{Short}$ models with different random seeds were trained using deepmd-kit[91,92] software. The $DP_{Short}$ models were trained with $4\times10^5$ steps. The embedding network had three layers with 25, 50 and 100 nodes and the fitting network was composed of three layers, each of which had 240 nodes. The max neighbor numbers were set as 90 for Li, 60 for C, 80 for H, 60 for O, 80 for P and 80 for F, which are sufficient for a 6 Å cutoff local environment descriptor. The Adam[93] method was used to minimize the loss functions with an exponentially decay learning rate from $1.00\times10^{-3}$ to $3.51\times10^{-8}$ used. In the exploration stage, we first performed the ConstQ, and then the ConstP MD simulations driven by the DP-QEq MLFF under the NVT ensemble (the temperature range was from the 200 K to 400 K and the simulation time was 20 ps) to extend the configurational space for achieving a sufficient sampling. We emphasize again that the training dataset expansion in the concurrent learning process is not driven by the ReaxFF simulations anymore, but instead by the targeted training potential – DP-QEq force field. This type of self-consistent strategy for optimizing MLFF has also been adopted by previous studies[94,95,96,97]. A single-interface model (548 atoms, the upper panel in **Figure 1c**) and a double-interface model containing a pair of counter electrodes (834 atoms, the lower panel in **Figure 1c**) were used in configurational sampling under the ConstQ and ConstP conditions, respectively. We fixed the uppermost/bottommost two layers of the top/bottom Li slab during those simulations. Other setups for performing MD simulations were consistent with those introduced in section 2.2.2. When the average force derivation (definitions can be found in Ref[90]) of a configuration was in the range 0.1 – 0.2 eV/Å, this configuration would be labelled by DFT calculations. After that, the QEq energy and forces (under a ConstQ constraint) were subtracted from the DFT results and the residuals were added to the next $DP_{Short}$ training iteration. As we discuss earlier, the physical meaning of the short-range term $E_{\mathrm{Short}}$ mainly refers to the local bonding interaction between atoms, which is less affected by the relatively long-range electrostatic environment. The impact of introducing bias potentials can be included in the Coulomb electrostatic interaction model of the QEq under a ConstP condition. In addition, traditional DFT is typically performed under a ConstQ constraint. Considering those two facts, the QEq energies and forces were always computed under the ConstQ constraint to generate the short-range interaction term $E_{Short}$ by Eq. 10 (shown in **Figure 1b**). When all average force derivations of all sampled configurations were lower than 0.1 eV/Å, the $DP_{Short}$ force field model can be regarded as converged. After finishing the concurrent learning iterations, we long-trained a $DP_{Short}$ model with $8\times10^6$ steps for subsequent simulations.

## Supporting Information

Details for the QEq energy expression; Charge equilibration under ConstQ and ConstP constraints; Solution of the QEq charge; Computation details of the electrostatic potential; Comparisons of the total energies and forces obtained from the DFT and full DP models on training dataset; Comparisons of the total energies and forces obtained from the DFT and full DP models on testing dataset; Comparisons of the short-range energies and forces obtained from DFT and DP-QEq $DP_{Short}$ models; Machine-learning force error analysis of the 4th round of ConstP testing trajectory; Decomposition of force errors into atomic types; Li charge comparisons; Final morphology of a 500 ps ConstQ MD simulation; Reaction mechanisms for various products; Details of the training dataset and testing dataset; Difference between the toy model and the realistic model; Li charge over time under different external potentials; Distributions of SEI inorganic components in the initial and final snapshots belonging to the 2nd round of ConstP DP-QEq simulation. ConstP testing simulations with a 3.3 Å cutoff for coordination number calculations; Changes of Li-Li bond numbers between the initial and the last electrochemical cycles; Computational workflow of our DP-QEq MD simulations.

## Acknowledgement

The authors gratefully acknowledge funding support from the Ministry of Science and Technology of the People's Republic of China (grant no. 2021YFB3800303), the National Natural Science Foundation of China (grant no. 52273223), DP Technology Corporation (grant no. 2021110016001141), the School of Materials Science and Engineering at Peking University, and the AI for Science Institute, Beijing (AISI). The computing resource of this work was provided by the Bohrium Cloud Platform (https://bohrium.dp.tech), which was supported by DP Technology. T. H and S. X sincerely acknowledge Dr. Jonas A. Finkler for inspiring discussions about QEq calculations.

## Code and Data Availability

The data supporting the reported findings, including all training dataset, force field model and complete scripts can be found in our github repo (https://github.com/sxu39/DP-QEq) and Bohrium platform (https://bohrium.dp.tech/notebooks/38175812597).

(2023).

# Supporting Information

## Observation of dendrite formation at Li metal-electrolyte interface: A machine-learning enhanced constant potential framework

Taiping Hu[1,2], Haichao Huang[3], Guobing Zhou[1,4], Xinyan Wang[5], Jiaxin Zhu[6], Zheng Cheng[2,7], Fangjia Fu[2,7], Xiaoxu Wang[5], Fuzhi Dai[2,8], Kuang Yu[*,3], Shenzhen Xu[*,1,2]

[1]Beijing Key Laboratory of Theory and Technology for Advanced Battery Materials, School of Materials Science and Engineering, Peking University, Beijing 100871, People's Republic of China

[2]AI for Science Institute, Beijing 100084, People's Republic of China

[3]Tsinghua-Berkeley Shenzhen Institute and Institute of Materials Research (iMR), Tsinghua Shenzhen International Graduate School, Tsinghua University, Shenzhen, 518055, People's Republic of China

[4]School of Chemical Engineering, Jiangxi Normal University, Nanchang 330022, People's Republic of China

[5]DP Technology, Beijing 100080, People's Republic of China

[6]State Key Laboratory of Physical Chemistry of Solid Surfaces, iChEM, College of Chemistry and Chemical Engineering, Xiamen University, Xiamen 361005, People's Republic of China

[7]School of Mathematical Sciences, Peking University, Beijing 100871, People's Republic of China

[8]School of Materials Science and Engineering, University of Science and Technology Beijing, Beijing, 100083, People's Republic of China

[*]Corresponding authors: yu.kuang@sz.tsinghua.edu.cn, xushenzhen@pku.edu.cn

## Contents

# 1. Methodology

## 1.1 Details for the QEq energy expression

The QEq energy $E_{\mathrm{QEq}}$ of a system is given by

$$E_{\mathrm{QEq}} = E_{\mathrm{Coulomb}} + E_{\mathrm{on-site}} = E_{\mathrm{Coulomb}} + \sum_{i=1}^{N} \left( \chi_i^0 Q_i + \frac{1}{2} J_i Q_i^2 \right) \quad \text{(S1)}$$

where $E_{\mathrm{Coulomb}}$ represents the Coulomb interactions of Gaussian charges. We first compute point charges interactions by the Ewald summation method[1] accelerated by the Particle Mesh Ewald (PME)[2] algorithm (Eq. S2), and then add a Gaussian charge distribution correction term (Eq. S6) to achieve an equivalent calculation of Gaussian charge Coulomb interactions[3,4,5]. $E_{\mathrm{on-site}}$ is the on-site energy. $Q_i$, $\chi_i^0$, and $J_i$ are the atomic charge, electronegativity and hardness of each atom in the modeling system, where the subscript "$i$" refers to atomic index. Atomic charges $Q_i$ will be computed by $E_{\mathrm{QEq}}$ minimization for every single configuration, and parameters of $\chi_i^0$ and $J_i$ of each element species are set up (see Table S1) at the beginning of molecular simulations.

Considering the periodic boundary condition in our simulations, the Coulomb interaction should be computed by the Ewald summation method[1]

$$E_{\mathrm{Ewald}} = E_{\mathrm{real}} + E_{\mathrm{recip}} + E_{\mathrm{self}} \quad \text{(S2)}$$

The real space part is

$$E_{\mathrm{real}} = \frac{1}{2} \sum_{i=1}^{N} \sum_{j \neq i}^{N_{\mathrm{neigh}}} Q_i Q_j \frac{\mathrm{erfc}\left( \frac{r_{ij}}{\sqrt{2}\eta} \right)}{r_{ij}} \quad \text{(S3)}$$

Here, $r_{ij}$ denotes the distance between atom $i$ and atom $j$, and $\eta$ is the width of the auxiliary charges. $N_{\mathrm{neigh}}$ indicates that the sum goes over all neighboring atoms within the real space cutoff radius $r_{\mathrm{cut}}$. For a specific $r_{\mathrm{cut}}$ and an error tolerance $\delta$, the parameter $\eta$ is given by $\sqrt{-\log 2\delta}/r_{\mathrm{cut}}$, which is a consistent setup as the OpenMM software[6].

The reciprocal space part is

$$E_{\text{recip}} = \frac{2\pi}{V} \sum_{\mathbf{k} \neq 0} \frac{\exp\left(-\frac{\eta^2 |\mathbf{k}|^2}{2}\right)}{|\mathbf{k}|^2} \left| \sum_{i=1}^{N} Q_i \exp(i\mathbf{k} \cdot \mathbf{r}_i) \right|^2 \tag{S4}$$

where $V$ is the volume of the unit cell and the sum goes over all reciprocal lattice points inside reciprocal space cutoff radius.

The self-interaction part is

$$E_{\text{self}} = -\sum_{i=1}^{N} \frac{Q_i^2}{\sqrt{2\pi}\eta} \tag{S5}$$

Traditional Ewald summation method scales with $O(N^2)$, we thus employ the $O(N \cdot \log N)$ scaling PME[2] method to calculate Coulomb interactions. In the PME calculations, the number of nodes in the mesh along each dimension is set as $2\kappa d/3\delta^{1/5}$, where $d$ is the width of the periodic box along the corresponding dimension and $\kappa$ is the Ewald splitting parameter.[6] The PME energy ($E_{\text{PME}} \equiv E_{\text{real}} + E_{\text{recip}} + E_{\text{self}}$) in this work is calculated by calling the DMFF software[7], an open-source automatic differentiable platform for molecular force field development.

Since we use the Gaussian charge distributions in the QEq calculation, the Gaussian charge correction term[4,5] should be added as:

$$E_{\text{corr}}^{\text{Gauss}} = -\frac{1}{2} \sum_{i=1}^{N} \sum_{j \neq i}^{N_{\text{neigh}}} Q_i Q_j \frac{\text{erfc}\left(\frac{r_{ij}}{\sqrt{2}\gamma_{ij}}\right)}{r_{ij}} + \sum_{i=1}^{N} \frac{Q_i^2}{2\sqrt{\pi}\sigma_i} \tag{S6}$$

$$\gamma_{ij} = \sqrt{\sigma_i^2 + \sigma_j^2} \tag{S7}$$

where $\sigma_i$ is the width of Gaussian charge density taken from the covalent radii of the relevant elements. We note that this correction term can be computed efficiently within a real space cutoff.

The presence of atomic charges may generate a large dipole, especially along the direction perpendicular to the interface (denoted as the $z$ direction in this work), the dipole correction[8] term should be added

$$E_{\mathrm{corr}}^{\mathrm{dipole}} = \frac{2\pi}{V}\left(M_z^2 - Q_{\mathrm{tot}}\sum_{i=1}^{N} Q_i z_i^2 - Q_{\mathrm{tot}}^2 \frac{L_z^2}{12}\right) \tag{S8}$$

$$M_z = \sum_{i=1}^{N} Q_i z_i \tag{S9}$$

$$Q_{\mathrm{tot}} = \sum_{i=1}^{N} Q_i \tag{S10}$$

where $L_z$ is the box length along the direction perpendicular to the electrode/electrolyte interface and $z_i$ denotes the $z$ component of $i$-th atomic coordinate. Because the total net charge has to be zero ($Q_{\mathrm{tot}} = 0$) as required by the neutrality of periodic supercells, $E_{\mathrm{corr}}^{\mathrm{dipole}}$ here only depends on the total dipole moment of the simulation cell.

Overall, the potential energy of a system can be written as

$$E_{\mathrm{QEq}} = E_{\mathrm{on-site}} + E_{\mathrm{PME}} + E_{\mathrm{corr}}^{\mathrm{Gauss}} + E_{\mathrm{corr}}^{\mathrm{dipole}} \tag{S11}$$

which will be used in subsequent charge equilibration calculations.

**Table S1.** Electronegativities and hardnesses parameters used in the QEq calculations. Those parameters are extracted from Ref[9].

| | Li | C | H | O | P | F |
|---|---|---|---|---|---|---|
| Electronegativity | -3.0000 | 5.8678 | 5.3200 | 8.5000 | 1.8000 | 9.0000 |
| Hardness | 10.0241 | 7.0000 | 7.4366 | 8.9989 | 7.0946 | 8.0000 |

## 1.2 Charge equilibration under ConstQ and ConstP constraints

For the QEq method[10] under a ConstQ constraint, the sum of all charges must be equal to the system total charge $Q_{\text{tot}}$ (Eq. S10), we thus employ the Lagrange multiplier method to solve this constrained minimization problem

$$\mathcal{L} = E_{\text{QEq}} - \chi_{\text{eq}}\left(\sum_{i=1}^{N} Q_i - Q_{\text{tot}}\right) \tag{S12}$$

The QEq charges $\{Q_i\}$ and the Lagrange multiplier $\chi_{\text{eq}}$ can thus be solved by

$$\begin{cases} \dfrac{\partial \mathcal{L}}{\partial Q_i} = 0 \\ \dfrac{\partial \mathcal{L}}{\chi_{\text{eq}}} = 0 \end{cases} \tag{S13}$$

resulting in a set of linear equations as follow:

$$\begin{bmatrix} A_{11} & \cdots & A_{1N} & 1 \\ \vdots & \vdots & \vdots & \vdots \\ A_{N1} & \cdots & A_{NN} & 1 \\ 1 & \cdots & 1 & 0 \end{bmatrix} \begin{bmatrix} Q_1 \\ \vdots \\ Q_N \\ \chi_{\text{eq}} \end{bmatrix} = \begin{bmatrix} -\chi_1^0 \\ \vdots \\ -\chi_N^0 \\ Q_{\text{tot}} \end{bmatrix} \tag{S14}$$

where $A_{ij}$ denotes the second-order derivative of the $E_{\text{QEq}}$ with respect to the atomic charges $Q_i$ and $Q_j$, i.e., $\partial^2 E_{\text{QEq}} / \partial Q_i \partial Q_j$, also known as the Hessian matrix, and is given by

$$A_{ij} = \begin{cases} J_i + \dfrac{1}{\sigma_i \sqrt{\pi}}, & \text{if } i = j \\ \dfrac{erf\left(\dfrac{r_{ij}}{\sqrt{2}\gamma_{ij}}\right)}{r_{ij}}, & \text{otherwise} \end{cases} \tag{S15}$$

For simplicity, we present here the expression for the *A* matrix elements in a non-periodic case. In fact, we use the project gradient algorithm that does not require the explicit construction of the *A* matrix (see Section 1.3).

In the ConstP condition, we apply external potentials to electrode atoms with a set of predefined values $\{\phi_i\}$, and the grand energy of this system is given by

$$\Omega = E_{\mathrm{QEq}} + \sum_{i=1}^{N} \phi_i Q_i \tag{S16}$$

$$\phi_i = \begin{cases} 0, & \text{if not Li} \\ 0, & \text{if Li, and } \mathrm{CN_{Li-Li}} < \mathrm{CN_{Metal}} \\ \phi_{\mathrm{Li,1}}, & \text{if Li, } \mathrm{CN_{Li-Li}} > \mathrm{CN_{Metal}}\text{, and Li} \in \text{Anode side} \\ \phi_{\mathrm{Li,2}}, & \text{if Li, } \mathrm{CN_{Li-Li}} > \mathrm{CN_{Metal}}\text{, and Li} \in \text{Cathode side} \end{cases} \tag{S17}$$

Here, $\mathrm{CN_{Li\text{-}Li}}$ and $\mathrm{CN_{Metal}}$ represent the coordination number (CN) only counting the Li-Li pair and the CN of an atom in the corresponding bulk phase (e.g. Li in this work), respectively. Our validations on the toy model show that an electrode electronegativity difference of 8 V can yield a relatively reasonable surface charge density distribution matching with an earlier theoretical study's result[11], we thus chose this potential difference. However, we note that owning to the presence of other elements (the electronegativities of which are fixed) in the Li metal-electrolyte interface model, not only the potential difference, but also the specific values of $\phi_{\mathrm{Li,1}}$ and $\phi_{\mathrm{Li,2}}$, should affect the QEq charge results in this realistic model. We tested three different $\{\phi_{\mathrm{Li,1}}, \phi_{\mathrm{Li,2}}\}$ sets, a symmetric set $\{\phi_{\mathrm{Li,1}} = -4\,V, \phi_{\mathrm{Li,2}} = +4\,V\}$, and two other sets of potential shifts relatively close to the above symmetric setup $\{\phi_{\mathrm{Li,1}} = -2\,V, \phi_{\mathrm{Li,2}} = +6\,V\}$ and $\{\phi_{\mathrm{Li,1}} = -6\,V, \phi_{\mathrm{Li,2}} = +2\,V\}$. We performed 100 ps ConstP MD simulations under different bias potentials, and counted the number of Li atoms deposited on the cathode side to represent the Li redox activity. We found that under the two setups of $\{\phi_{\mathrm{Li,1}} = -4\,V, \phi_{\mathrm{Li,2}} = +4\,V\}$ and $\{\phi_{\mathrm{Li,1}} = -6\,V, \phi_{\mathrm{Li,2}} = +2\,V\}$, no effective Li depositions were observed, while for the setup of $\{\phi_{\mathrm{Li,1}} = -2\,V, \phi_{\mathrm{Li,2}} = +6\,V\}$, the number of Li atoms at the cathode side increased by 93 (~ 2 layers of Li deposition) along this 100 ps ContP MD trajectory. We thus employed the set of $\phi_{\mathrm{Li,1}} = -2\,\mathrm{V}$ for the anode side and $\phi_{\mathrm{Li,2}} = +6\,\mathrm{V}$ for the cathode side in our simulations. We also acknowledge that our potential setup values deviate from the actual operating voltages of Li-ion batteries. It would be a direction for the future development of our method to optimize the parameters of electronegativity shifts. However, we emphasize that although the machine learning force field (MLFF) offers a significantly improved computational efficiency, the time scale of MD simulations remains relatively short compared to the real-world scenarios. We thus need to employ a relatively large bias potential to enhance the driving force of electrochemical reactions. The $\{\phi_{\mathrm{Li},1} = -2\ \mathrm{V},\ \phi_{\mathrm{Li},2} = +6\ \mathrm{V}\}$ parameter setup in our current study is able to achieve a practical electrochemical simulation for observing the dynamic process of Li dendrite formation, we thus used this setup throughout the work.

In principle, an open system under a ConstP condition allows for the presence of a net charge in simulations. However, the whole cell in our study must maintain neutrality due to the periodic boundary condition, we thus add the neutral constraint ($Q_{\mathrm{tot}} = 0$) in the ConstP condition. The Lagrangian should be written as

$$\mathcal{L} = \Omega - \chi_{\mathrm{eq}} \left( \sum_{i=1}^{N} Q_i - Q_{\mathrm{tot}} \right) \tag{S18}$$

The QEq charges $\{Q_i\}$ and the Lagrange multiplier $\chi_{\mathrm{eq}}$ can also be solved by using Eq. S13. This also results in a set of linear equations as follow:

$$\begin{bmatrix} A_{11} & \dots & A_{1N} & 1 \\ \vdots & \vdots & \vdots & \vdots \\ A_{N1} & \cdots & A_{NN} & 1 \\ 1 & \cdots & 1 & 0 \end{bmatrix} \begin{bmatrix} Q_1 \\ \vdots \\ Q_N \\ \chi_{\mathrm{eq}} \end{bmatrix} = \begin{bmatrix} -\chi_1^0 + \phi_1 \\ \vdots \\ -\chi_N^0 + \phi_N \\ Q_{\mathrm{tot}} \end{bmatrix} \tag{S19}$$

## 1.3 Solution of the QEq charge

Directly solving Eq. S14 and Eq. S19 is computationally expensive, which mainly originates from the explicit construction of the *A* matrix and the solution of its inversion, thus hindering broader applications of the QEq method to larger systems. The original reactive force field (ReaxFF)[12,13,14,15] combined with the QEq method treats Coulomb interactions up to a short cutoff for better computational performance. However, a recent work by Nwankwo et al[16] showed that the Ewald summation method offers a more accurate representation of Coulomb interactions in some systems, such as capacitors and charged electrodes. Consequently, developing an efficient approach to solve Eq. S14 and Eq. S19 within the Ewald summation framework is of great importance.

In our work, we first employ the projection gradient algorithm combined with the automatic differentiation technique of the *JAX* library[17] and then use LBFGS[18] algorithm in the *jaxopt* library[19] to resolve this constrained minimization problem. This solution exhibits two significant advantages. First, it only requires the gradient information of the $E_{\mathrm{QEq}}$ and avoids explicit construction of the Hessian matrix, which can remarkably reduce the use of memory, making it applicable to larger systems. Second, thanks to the automatic differentiation technique, we can quickly obtain the gradient with just few lines of code, avoiding complex analytical derivations. Moreover, the *JAX* library's just-in-time (JIT) compilation feature significantly improves computational speed on graphics processing unit (GPU) hardware. Table S2 presents the computational time as a function of the total atom number in the system. All calculations were performed on the NVIDIA V100 GPU device. We can see that our current implementation shows a quasi-linear scaling, especially for large systems.

The LBFGS algorithm iteratively solves QEq charges and a good initial guess can accelerate the convergence of iterations. Therefore, at the very beginning of our MD trajectory (i.e. the first MD step), we employ the matrix inversion method to compute QEq charges, and then use this result as the initial guess for the projection gradient method solving charges at the second MD step. Similarly, we use the charge solution of the second step as the initial guess for the third MD step, and so on and so forth. In one word, for application of the projection gradient algorithm, we always read QEq charges from the previous step as the initial guess for the subsequent step, which helps us achieve a good numerical stability of LBFGS algorithm.

**Table S2.** Time costs of QEq calculations as a function of total atom numbers in our modeling systems.

| Total atomic numbers | 500 | 1000 | 2000 | 4000 | 8000 | 16000 |
|---|---|---|---|---|---|---|
| Time costs (s) | 0.029 | 0.032 | 0.037 | 0.077 | 0.190 | 0.413 |

## 1.4 Computation details of the electrostatic potential

We employ the probing charge method (formula shown as Eq. S20), equivalent to solving Poisson's equation, to calculate the electrostatic potential

$$\phi = \frac{E_{\mathrm{Coulomb}}(\{Q_i\} + q_{\mathrm{probing}}) - E_{\mathrm{Coulomb}}(\{Q_i\})}{q_{\mathrm{probing}}} \quad \text{(S20)}$$

where $E_{\mathrm{Coulomb}}(\{Q_i\} + q_{\mathrm{probing}})$ and $E_{\mathrm{Coulomb}}(\{Q_i\})$ represent the electrostatic energies ($E_{\mathrm{PME}} + E_{\mathrm{corr}}^{\mathrm{Gauss}} + E_{\mathrm{corr}}^{\mathrm{dipole}}$) of the systems with and without the probing charge, respectively. $q_{\mathrm{probing}}$ is the magnitude of the probe charge and set as $10^{-5}$ e.

# 2. Supplementary Figures and Tables

## 2.1 Comparisons of the total energies and forces obtained from the DFT and full DP models on training dataset

We train a full DP model using the dataset calculated by the DFT method. The model's performance on the **training dataset** is displayed in Figure S1-S3.

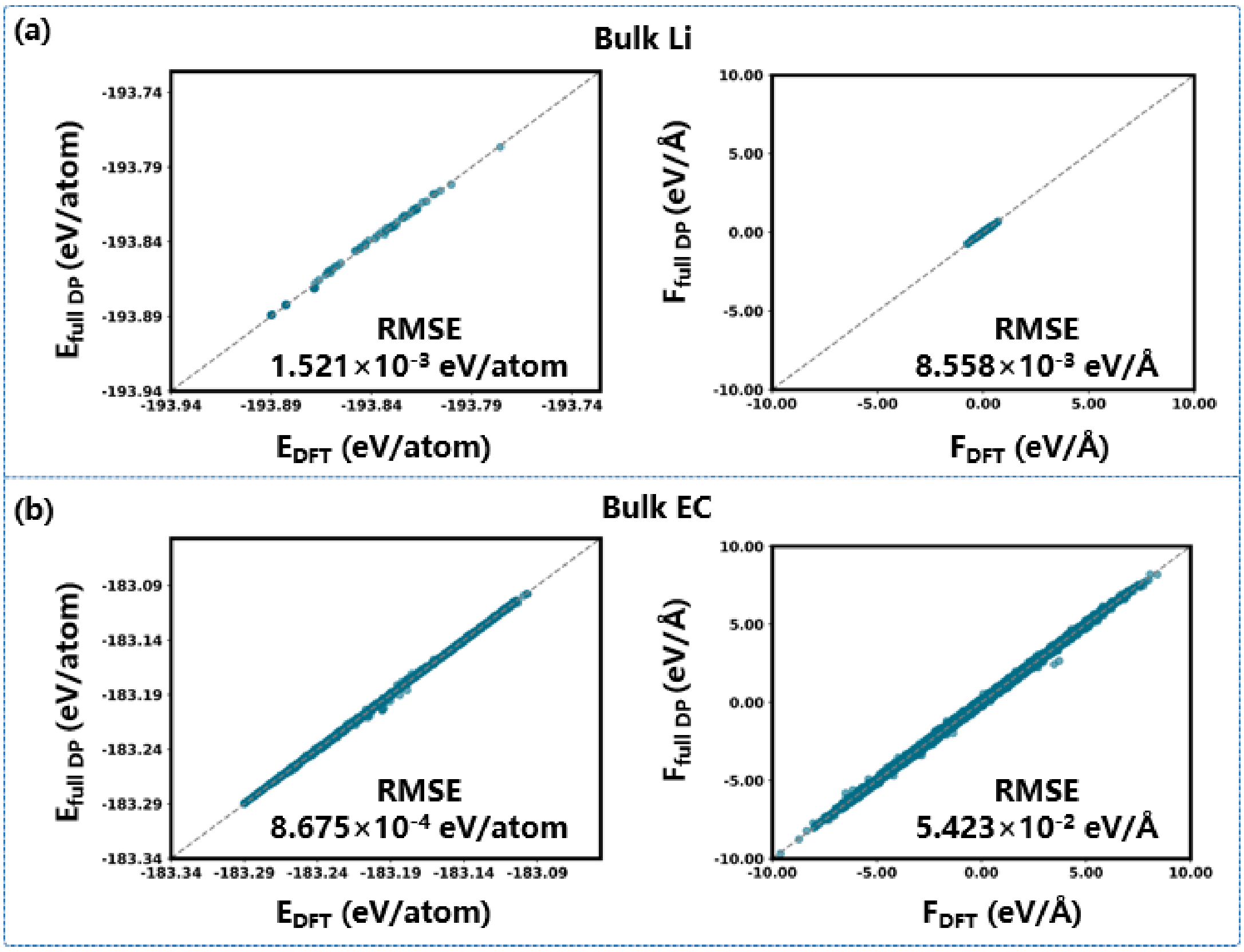

**Figure S1.** Comparisons of the total energies and forces obtained from the DFT and full DP models for (a) the bulk Li and (b) the bulk EC on the training dataset.

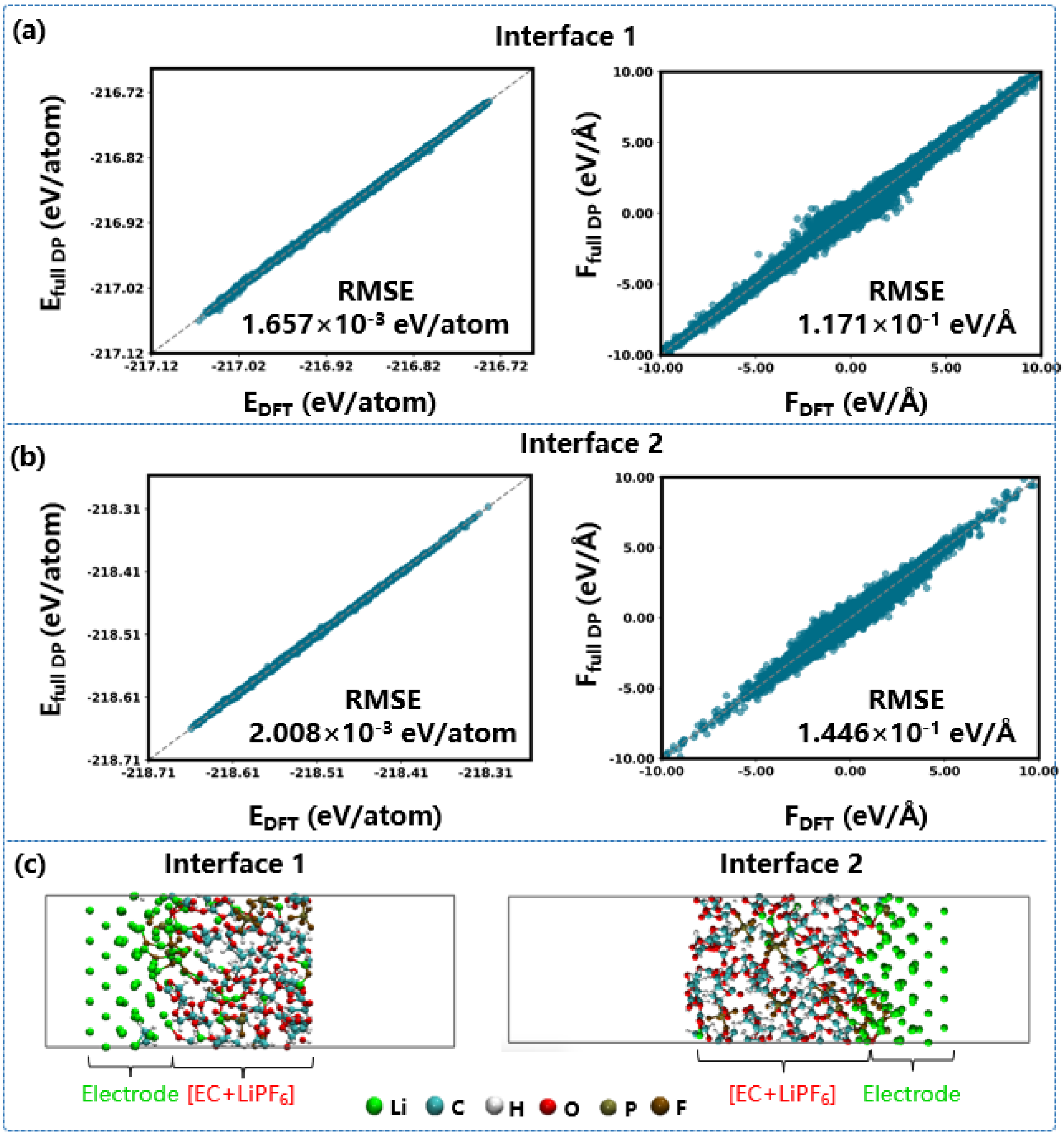

**Figure S2.** Comparisons of the total energies and forces obtained from the DFT and full DP models for (a) the interface 1 and (b) the interface 2 models on the training dataset. (c) Atomic structures of the corresponding interface models used in the dataset construction. Visualizations are done in the VMD[20] software. Interface 1 and 2 models have different numbers of atoms.

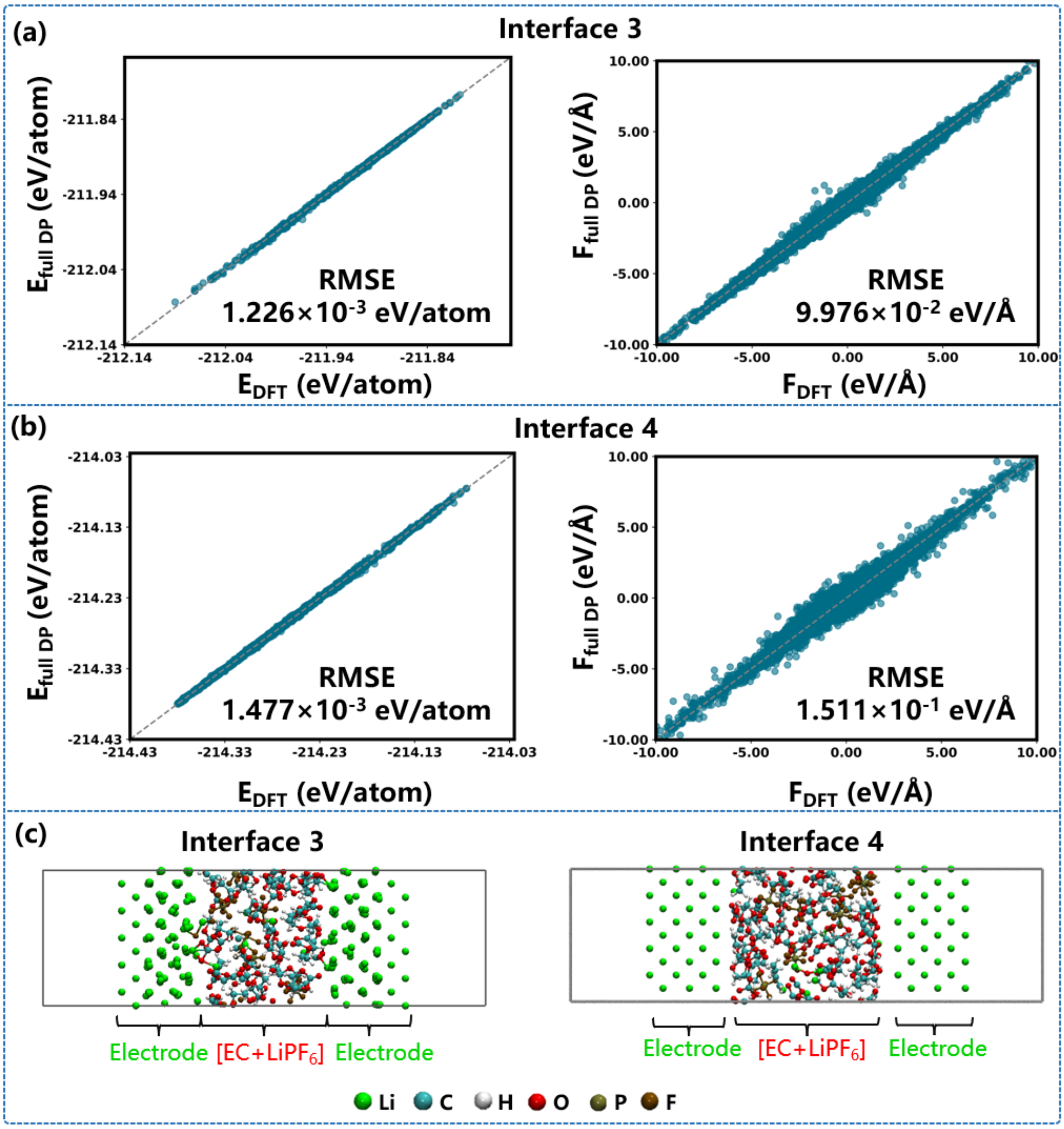

**Figure S3.** Comparisons of the total energies and forces obtained from the DFT and full DP models for (a) the interface 3 and (b) interface 4 models on the training dataset. (c) Atomic structures of the corresponding interface models used in the dataset construction. Visualizations are done in the VMD[20] software. Interface 3 and 4 models have different nusmbers of atoms.

## 2.2 Comparisons of the total energies and forces obtained from the DFT and full DP models on testing dataset

The full DP model's performance on the **testing dataset** is displayed in Figure S4. We performed 100 ps NVT simulations at 300 K for the bulk Li, bulk [EC+$LiPF_6$] and Li/[EC+$LiPF_6$] interface models, and extracted configurations along the trajectories at 1 ps interval for DFT calculations. This simulation time significantly exceeds that used in the exploration stage (20 ps) during the DPGEN iterations and these DFT points can thus be regarded as the testing dataset. Those results can validate the good performance of the full DP model.

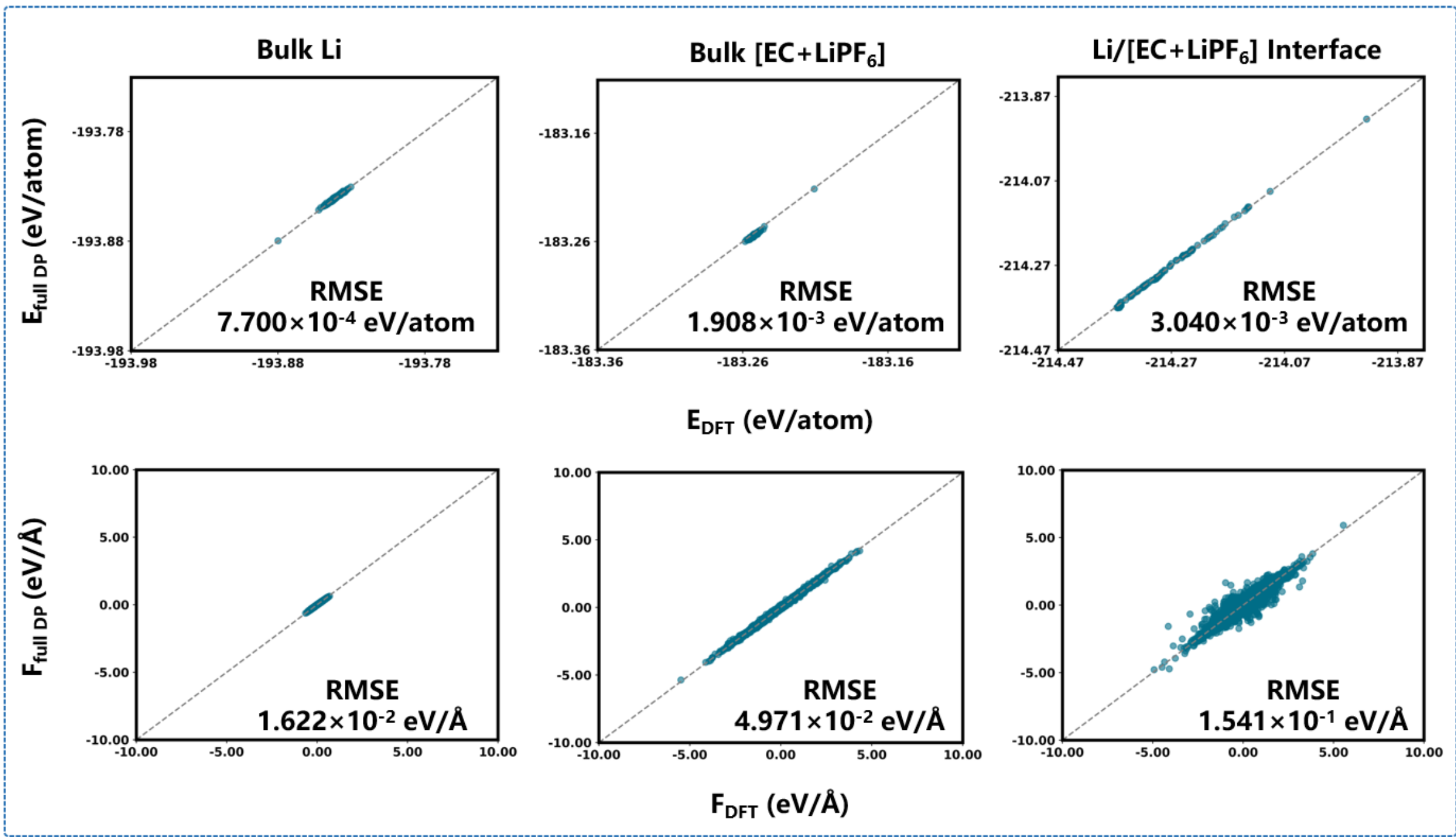

**Figure S4.** Comparisons of the total energies and forces obtained from the DFT and full DP models for the bulk Li, bulk [EC+$LiPF_6$] and Li/[EC+$LiPF_6$] interface.

## 2.3 Comparisons of the short-range energies and forces obtained from DFT and DP-QEq $DP_{Short}$ models

We also trained a $DP_{Short}$ model by subtracting the long-range QEq part from the DFT total energy and forces results. The model's performance on the **training dataset** is displayed in Figure S5-S7. Corresponding results on the testing dataset are shown in main text. Those results validate the reliability of our $DP_{short}$ model.

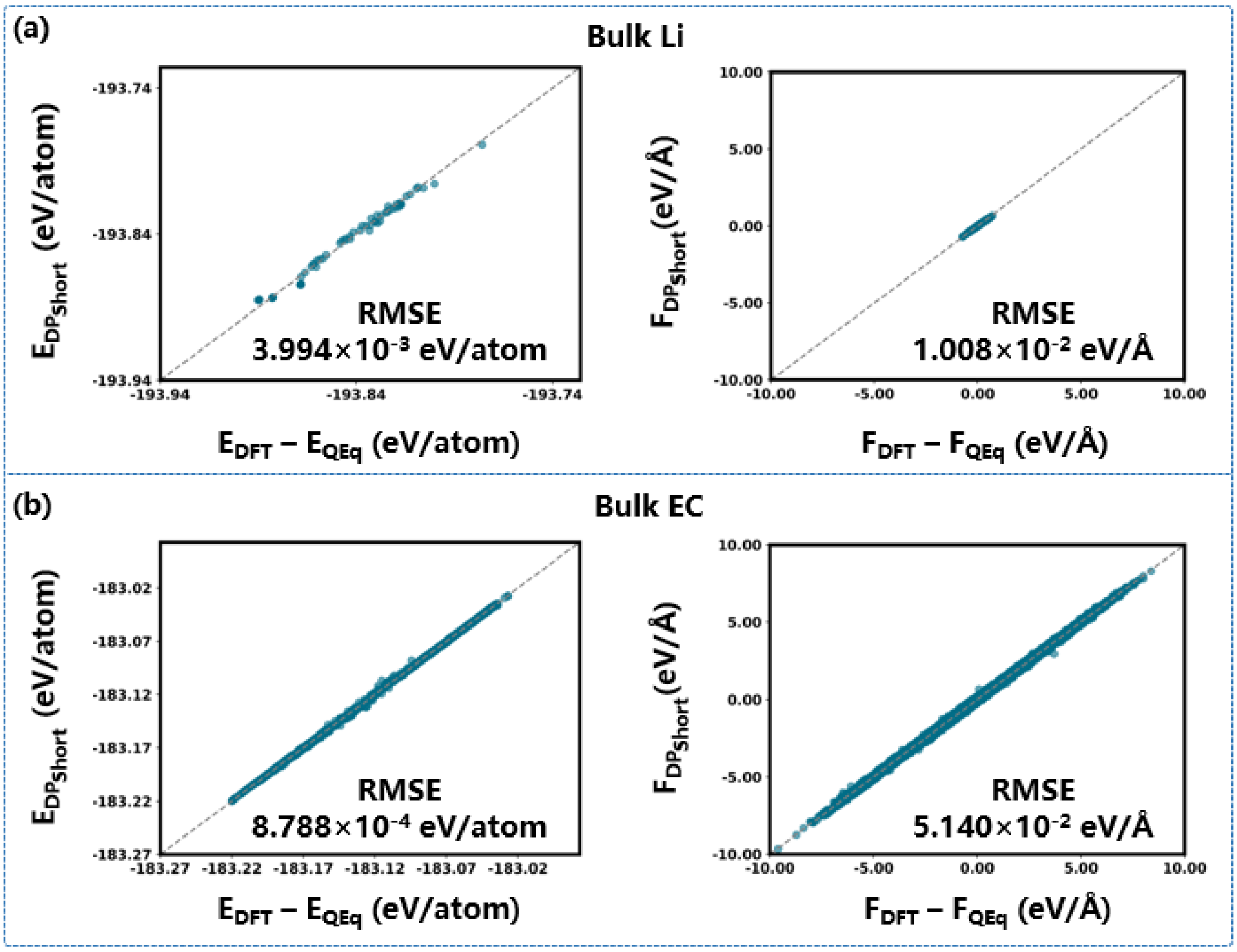

**Figure S5.** Comparisons of the short-range energies and forces obtained from DFT and DP-QEq models on the training dataset for (a) the bulk Li and (b) the bulk EC systems.

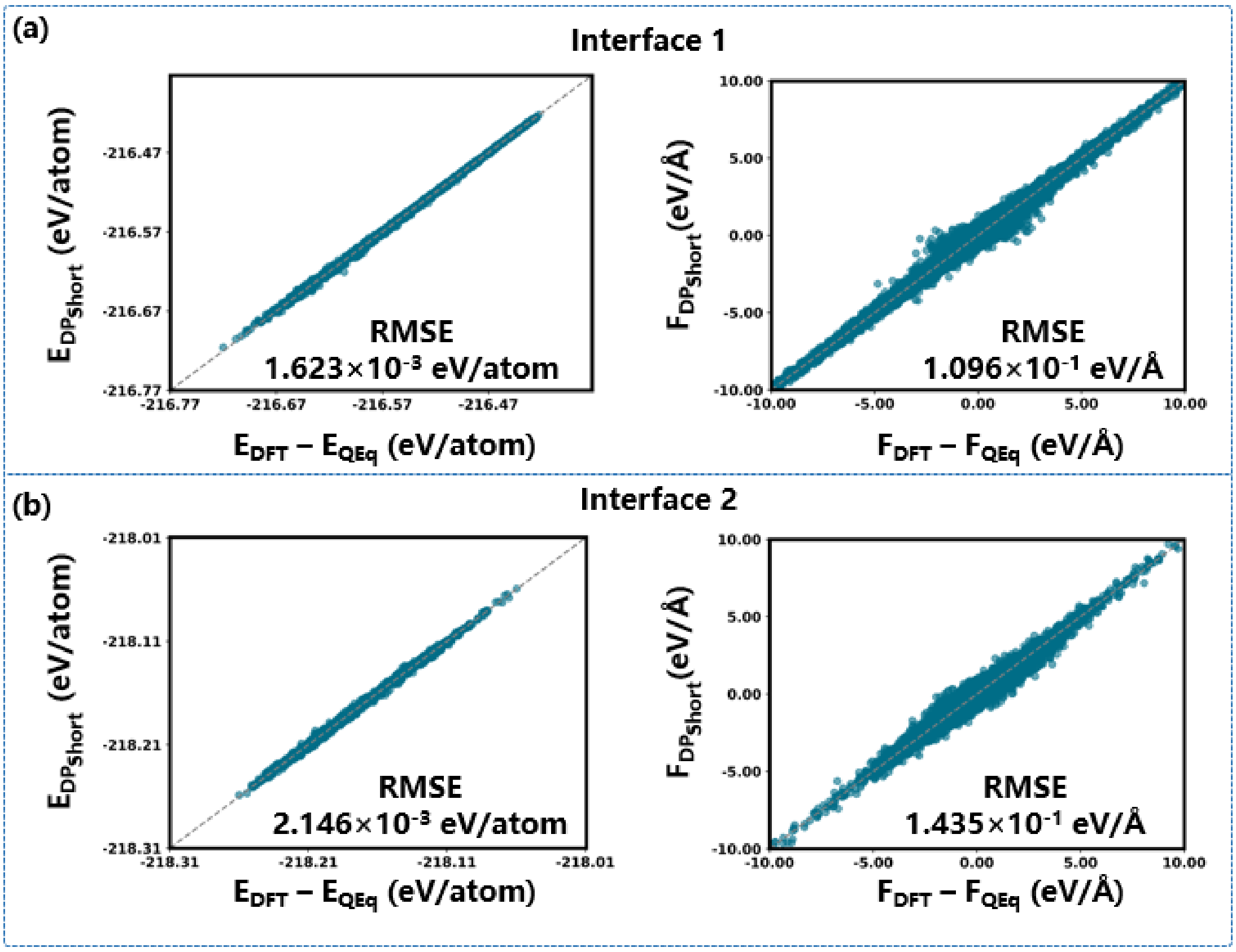

**Figure S6.** Comparisons of the short-range energies and forces obtained from DFT and DP-QEq models on the training dataset for (a) the interface 1 and (b) the interface 2 systems (the corresponding atomic structures presented in Figure S2).

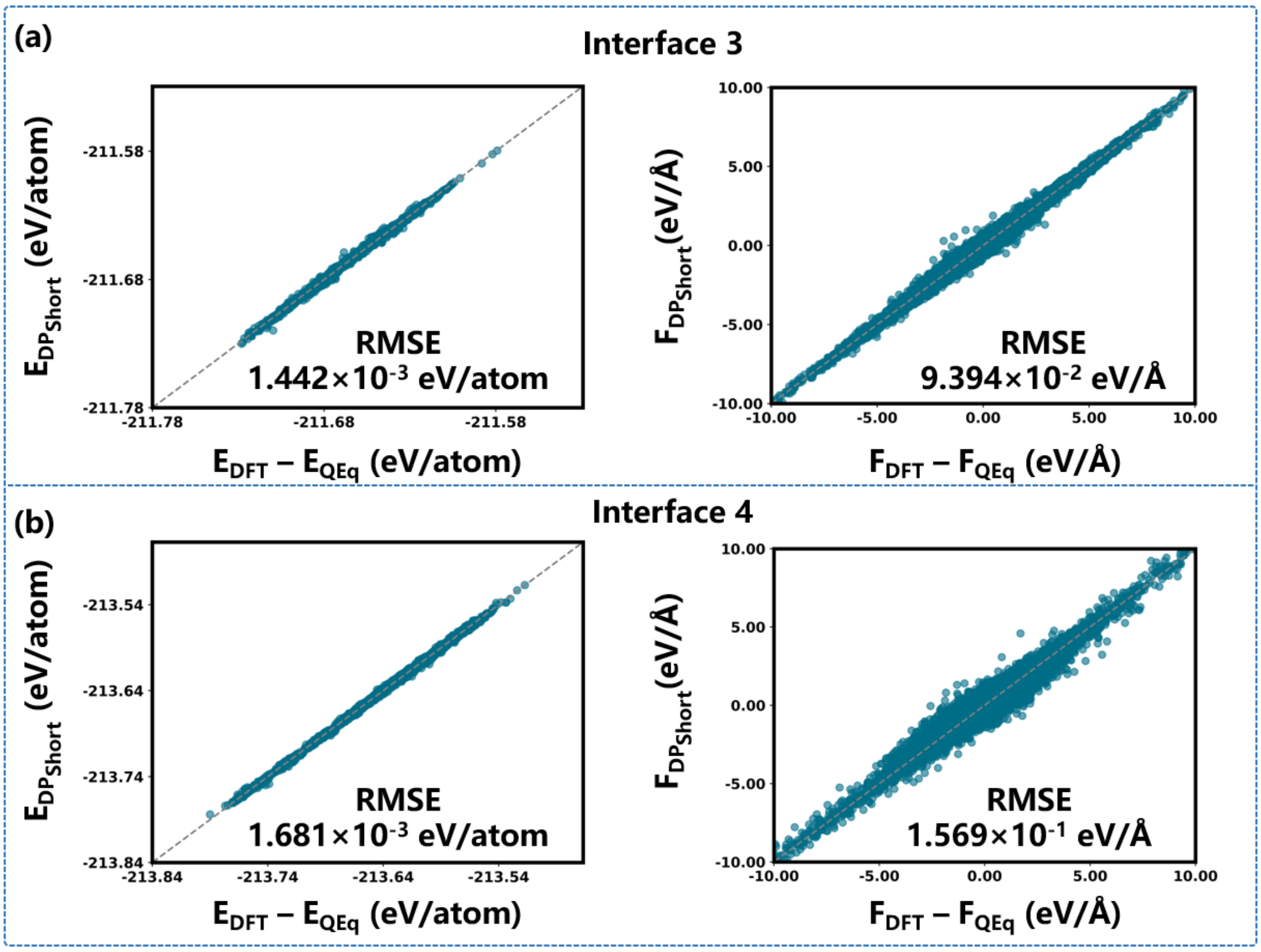

**Figure S7.** Comparisons of the short-range energies and forces obtained from DFT and DP-QEq models on the training dataset for (a) the interface 3 and (b) the interface 4 systems (the corresponding atomic structures presented in Figure S3).

## 2.4 Machine-learning force error analysis of the 4th round of ConstP testing trajectory

In the main text, we showcase the overall force error between DP-QEq and DFT models on the testing dataset, which consists of a 50 ps ConstQ and 4 rounds of 100 ps ConstQ trajectories. To further specify the force error origin, especially to check out the atomic force errors on Li metal slab and the nucleated Li dendrite which are the most relevant regions to our scientific topic, we first identify metallic Li atoms (including the Li electrode and the emerged Li dendrite atoms) within the SEI along the 4th round of ConstP trajectory and track the force errors of those Li atoms, in terms of RMSE compared to DFT-predicted forces. As shown in Figure S8, the force errors of these Li atoms involved in Li dendrite nucleation are quite small (around 0.07 – 0.09 eV/Å) throughout the entire testing trajectory, indicating results of the dynamic process of Li dendrite formation observed in our work are physically meaningful.

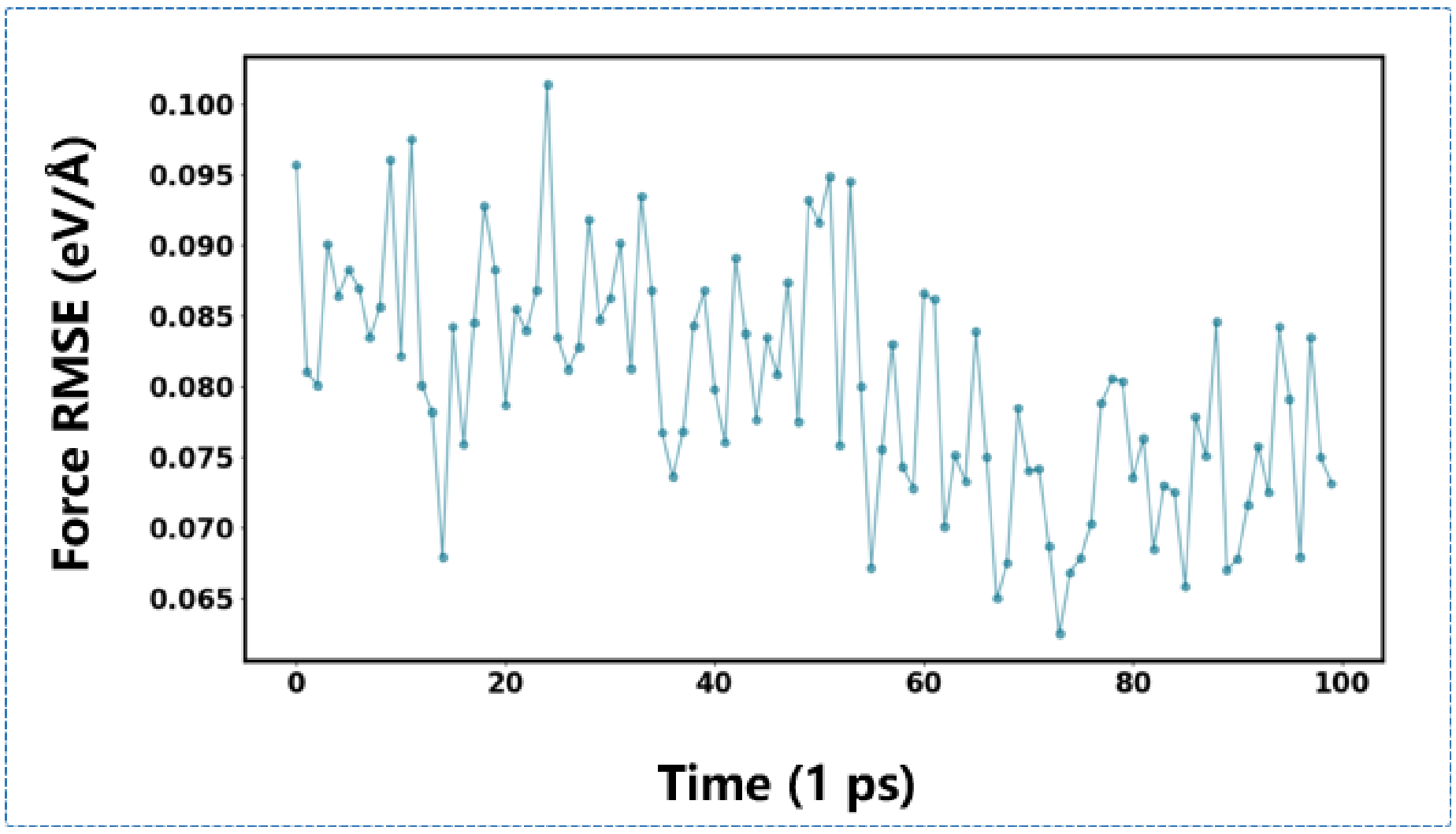

**Figure S8.** RMSE values of atomic forces between our DP-QEq and DFT methods as a function of the timestep during the 4th round of ConstP testing trajectory. Only metallic Li atoms in our interfacial model (including the emerged Li dendrite in SEI) are taken into consideration.

## 2.5 Decomposition of force errors into atomic types

In the main text, we label the potential energies for 200 configurations along a 500 ps ConstQ trajectory (shown in **Figure 2c**). In this section, we first estimate the overall force RMSEs (full DP vs DFT and DP-QEq vs DFT) as a function of timestep in Figure S9a, and find that the RMSEs both in the full DP and DP-QEq models remain consistent with the results shown in **Figure 2a**. We further decompose the force errors into each atomic type (see Figure S9b and S9c), and find that the maximum force errors appear on the P atoms. The relatively small force errors on Li atoms, which are most relevant to our investigated topic "Li dendrite formation", further indicate that our simulated results are meaningful.

We can also see that the force RMSE of P atoms is relatively large. After carefully examining all simulation trajectories, we find the F ions in $PF_6^-$ tend to gradually dissociate from the $PF_6^-$ anion and form LiF with Li ions, which is the well-acknowledged mechanism for the LiF formation. The resulting challenge is that P atoms would experience more complicated variation of coordination environment along our simulations (changing from a 6-fold coordination with F anions to a single P ion), that's why the P atoms are more difficult to be accurately described by the MLFF and exhibit a larger force deviation. However, we emphasize that this does not affect the validation of our proposed DP-QEq algorithm, and also has little impact on the major conclusions of our simulation results, because P atoms neither appear in the major components of the SEI layer ($Li_2O$, LiF, and $Li_2CO_3$), nor participate in the initial nucleation of Li dendrites. To demonstrate this, we calculate the accumulative coordination numbers (CN) of the Li-O, Li-F and Li-P pairs in the 4th round DP-QEq ConstP simulation (see Figure S9d). Here, only the Li atoms in the dendrite nucleation region are selected for the CN calculations. A total of 1500 snapshots, taken at 0.2 ps intervals along the whole trajectory, are used for statistical computation of CN values. We can see obviously that, the small Li-P CN values in Figure S9d indicate the Li atoms in the dendrite nucleation region almost do not coordinate with P atoms, thus justifying our argument discussed above.

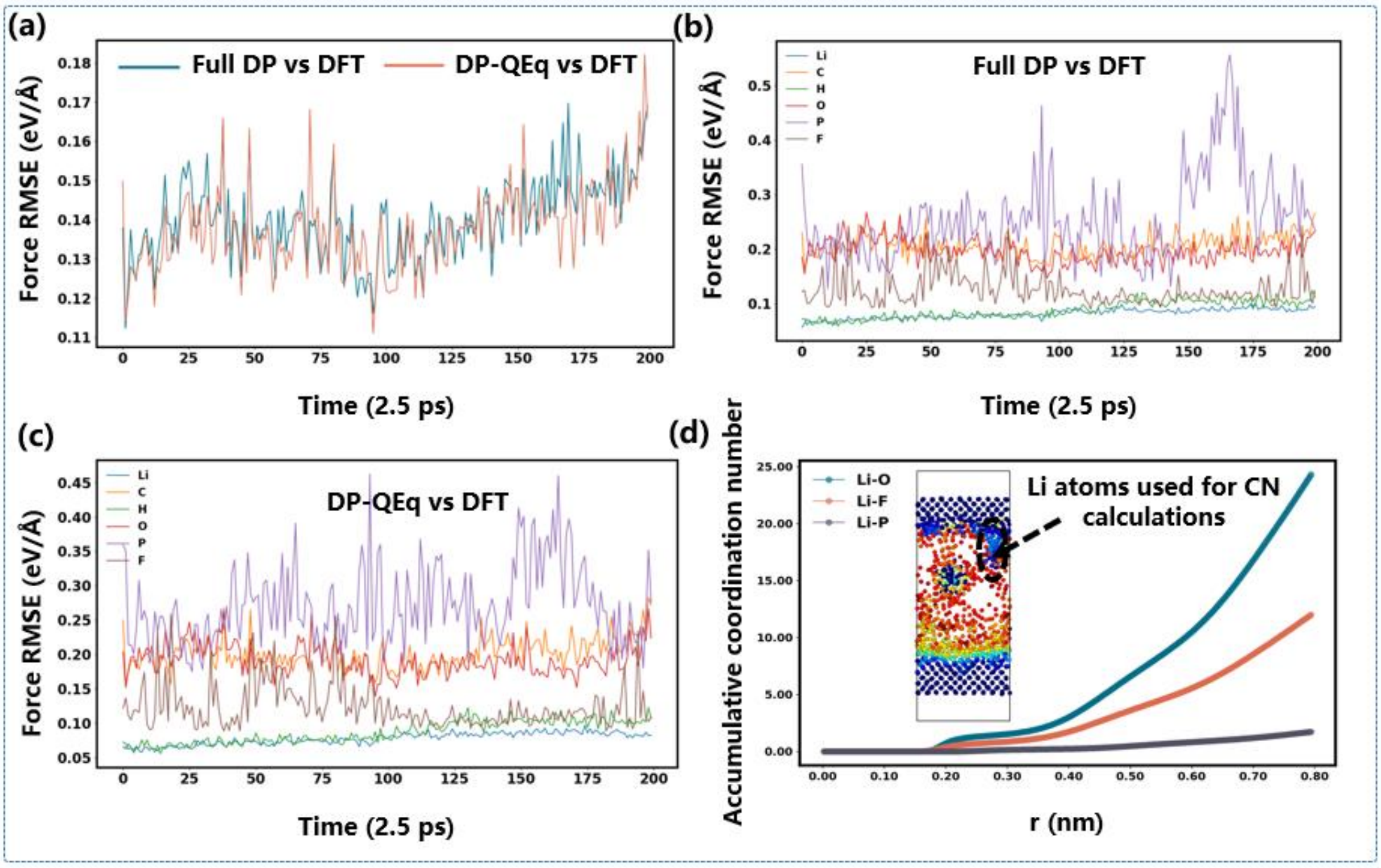

**Figure S9.** (a) Force RMSEs (Full DP vs DFT, and DP-QEq vs DFT) as a function of timestep along a 500 ps ConstQ trajectory. The configurations used for force comparisons are selected from the same trajectory driven by the ConstQ DP-QEq model. The RMSEs between (b) full DP vs DFT and (c) DP-QEq vs DFT are further decomposed into each atomic type. (d) Accumulative coordination numbers of the Li-O, Li-F and Li-P pairs in the $4^{th}$ round DP-QEq ConstP simulation. Only the Li atoms in the dendrite nucleation region are selected for the coordination number calculations.

## 2.6 Li charge comparisons

In this section, we first compare the QEq charges with those from the *ab-initio* DFT method. Here we consider several different post-processing charge analysis schemes, including the Hirshfeld[21], CM5[22] and Bader[23,24] models (Figure S10a). We find that: (1) there exists discrepancies among different post-processing charge analysis schemes; (2) the Bader charge model yields unphysical predictions, as unexpected non-zero net charges emerge within the interior region of metallic electrodes; (3) both the Hirshfeld and CM5 charge analysis schemes provide qualitatively reasonable charge results based on DFT calculations, given that charges within the Li metal interior region are predicted to be ~ 0 by Hirshfeld and CM5 schemes; (4) the charge distribution predicted by the QEq method qualitatively match well with the *ab-initio* charge results generated by Hirshfeld and CM5 methods.

We then perform DFT calculations with an extra electron added into the system ($\Delta N_e = 1$). Here $\Delta N_e = 1$ means that the total electron number in DFT calculations increases by one, which is a common feature in most DFT packages. Please note that adding more electrons does not indicate that the system exhibits net non-zero charge, we have to implement compensating charge into the cell due to the neutrality constraint required by the periodic boundary condition. This function, including an adjustable total number of electrons with a compensating charge plate whose position is also configurable (usually placed in the vacuum region above the electrolyte layer), is implemented in the ABACUS[25,26] code. The dipole correction is also included in these calculations. The purpose of introducing an extra electron is to mimic a stronger reducing condition. We then calculate the Hirshfeld and CM5 atomic charges and plot their distributions (Figure S10b). We can see that introducing an extra electron does not lead to the emergence of negative charges on metallic Li atoms in the electrode. These results can demonstrate that the presence of $Li^+$ ions in the electrolyte environment near electrode surfaces can accommodate negative charges in the cathode side, thus a reduction reaction of $Li^+ \rightarrow Li^0$ occurs when a reducing potential is applied in a ConstP condition.

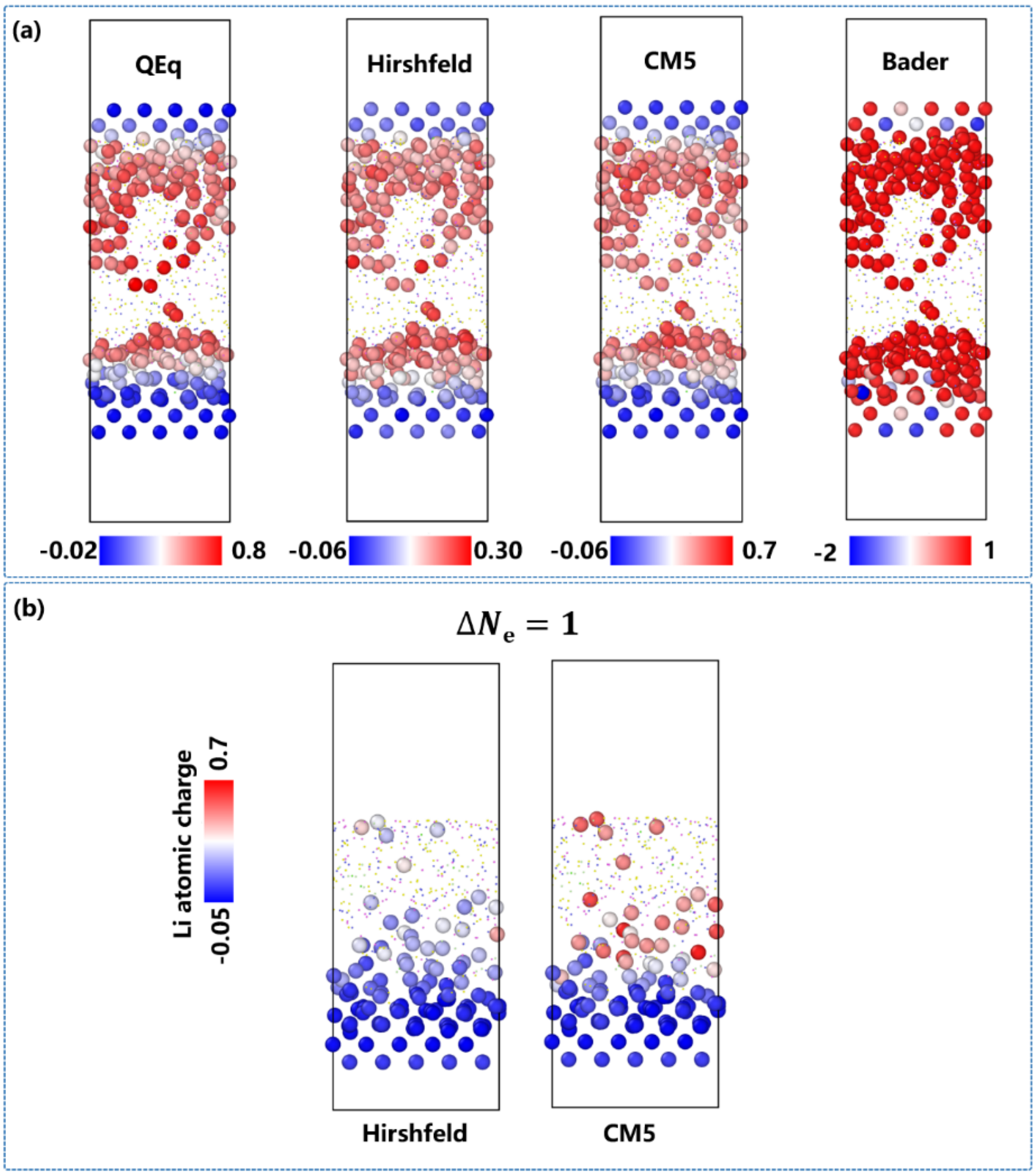

**Figure S10.** (a) Li atomic charges calculated by the QEq method, and by the charge analysis approaches of Hirshfeld[21], CM5[22] and Bader[23,24] based on DFT calculations for the Li/[EC+$LiPF_6$] interfacial model shown in the lower panel of Figure 1c in the main text. (b) Li atomic charge distribution (calculated by the Hirshfeld and CM5 methods) obtained from DFT calculations, with ($\Delta N_e = 1$) an extra electron added into the interfacial system.

## 2.7 Final morphology of a 500 ps ConstQ MD simulation

In the main text, we validate our DP-QEq method under ConstQ and ConstP conditions. The supporting results for the ConstQ validations are displayed in Figure S11. More details can be found in the main text. In this part, we first conduct a 200 ps ConstQ NPT simulation to relax to avoid cavity generations caused by interfacial side reactions between the Li metal and electrolytes. We then perform a 500 ps ConstQ NVT MD to generate a configuration used for subsequent ConstP simulation. We note here that the current simulation time of a MD trajectory cannot guarantee the system fully reaching a equilibrium state with all possible surface reactions completely occuring. However, we just focus on capturing dynamic evolutions of the Li/[EC+Li$PF_6$] interface in this work, which are inherently non-equilibrium processes in realistic batteries. It is thus unnecessary to extend our simulation time untill all possible SEI products' contents reaching exact plateaus with respect to time, which is also impractical considering the computational cost.

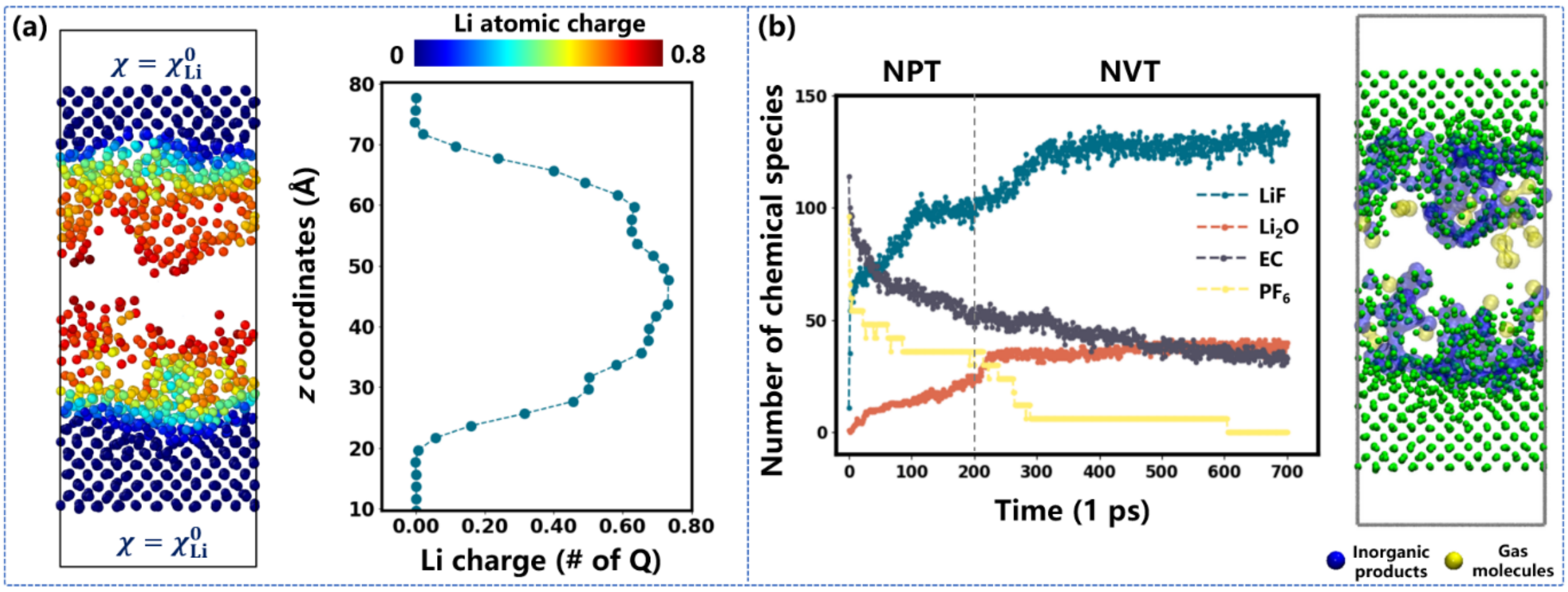

**Figure S11.** (a) The final morphology obtained from a 500 ps ConstQ MD simulation (left). We only display the Li atoms for clarity, which are color-coded by their atomic charges. The visualization is performed by the OVITO software[27]. The Li atomic charge distribution curve as a function of *z* coordinates is shown by the side (right). (b) Molecule numbers of various chemical species as a function of the MD time steps (left) and the spatial distributions of various products (left) in the double-electrode model. The visualization is done by the VMD software.[20]

## 2.8 Reaction mechanisms for various products

In our simulations, the EC molecule initially undergoes a ring-opening reaction to form $CO_3^{2-}$ and $C_2H_4$, with $CO_3^{2-}$ further decomposed to $CO_2$, followed by a subsequent decomposition step to CO. The F ions in $PF_6^-$ can dissociate and combine with Li ions to form LiF. We plot the formation mechanisms of these products in Figure S12.

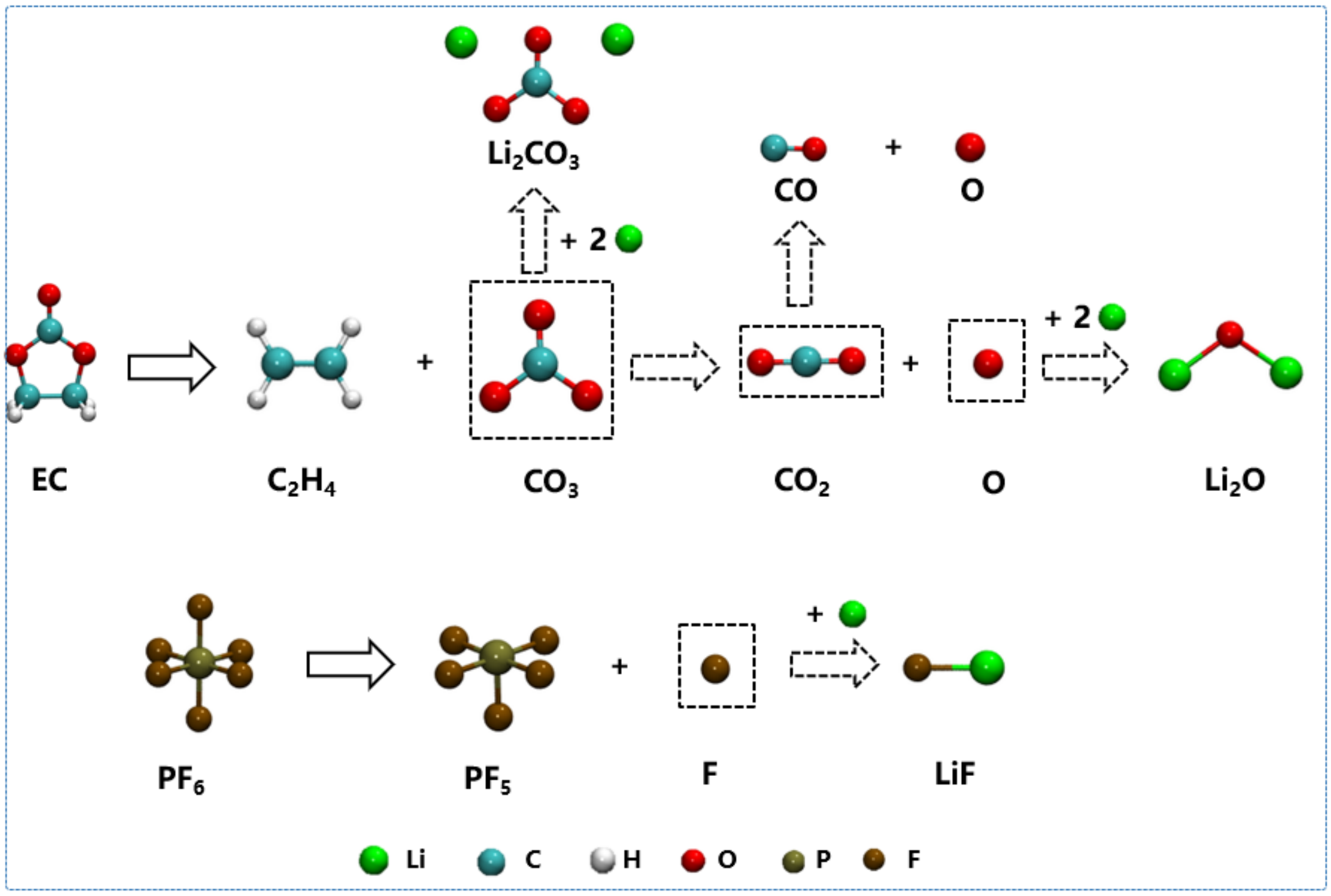

**Figure S12.** Formation mechanisms of the SEI inorganic components ($Li_2CO_3$, $Li_2O$ and LiF) and gases ($C_2H_4$, $CO_2$ and CO).

## 2.9 Details of the training dataset and testing dataset

We list the amounts of various decomposition products included both in the training dataset and testing dataset in **Table S3**. We also use the ReacNetGenerator program[28] (same with that mentioned in the main text) to identify these chemical species in our atomic model. Comparisons of the forces projected on each of those decomposition products between the DFT results and our DP-QEq MLFF predictions in the final snapshot of the testing dataset are displayed in Figure S13. We can see that the force deviations of those products are around the same quantitative level with the overall force deviation in the complete testing dataset (see Figure 2a), justifying the validity of our trained MLFF model in terms of describing those decomposition products.

**Table S3.** Numbers of various decomposition product species included in the training dataset and testing dataset.

| | $Li_2O$ | LiF | $Li_2CO_3$ | CO | $CO_2$ | $C_2H_4$ |
|---|---|---|---|---|---|---|
| Training dataset | 24469 | 162154 | 12979 | 5006 | 2201 | 32327 |
| Testing dataset | 5419 | 16795 | 666 | 1702 | 564 | 2223 |

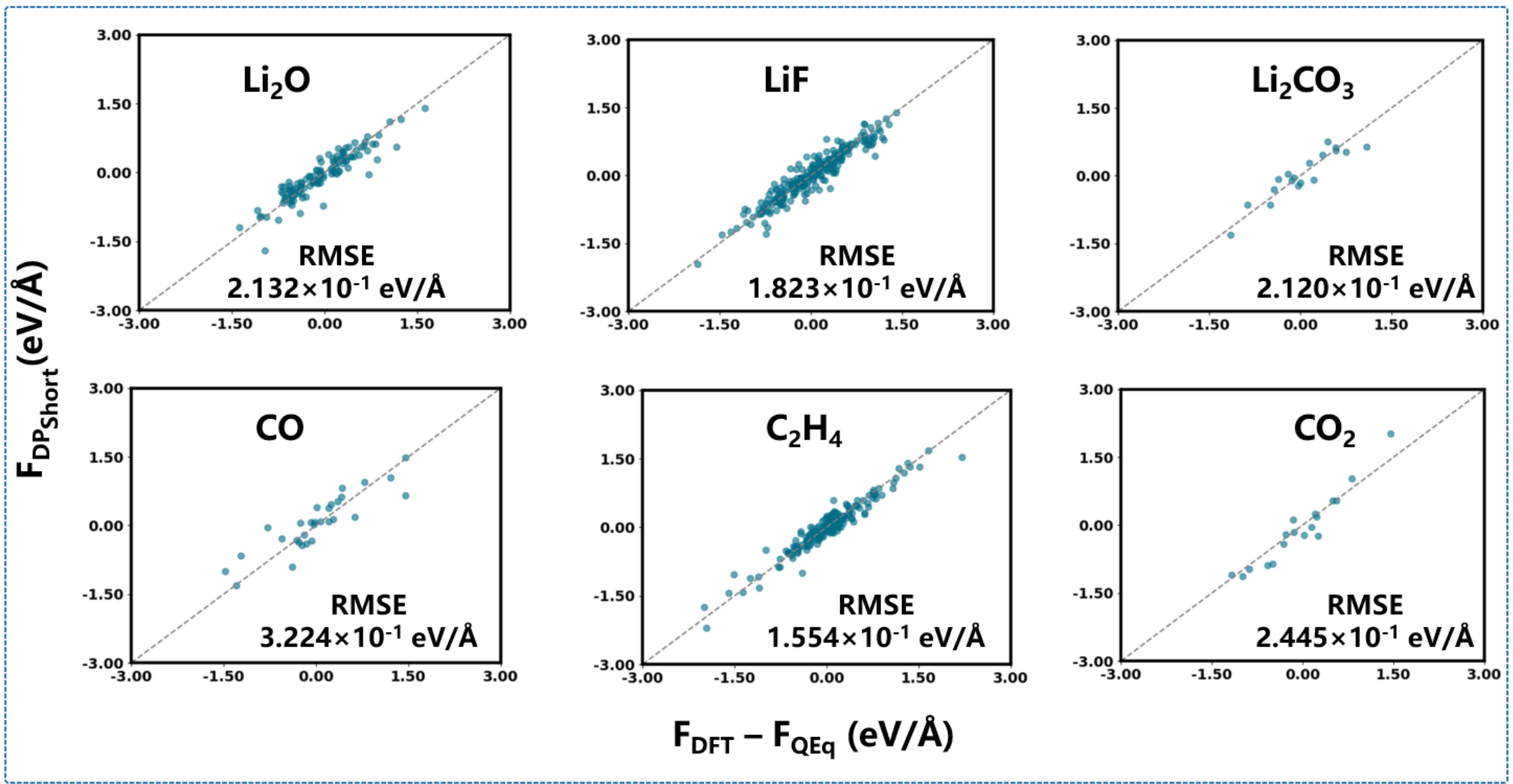

**Figure S13.** Comparisons of forces projected on each of decomposition products between the DFT results and our DP-QEq MLFF predictions in the final snapshot of the testing dataset.

## 2.10 Difference between the toy model and the realistic model

In the main text, we validate the DP-QEq method by a toy model and a realistic model. Although both models are Li metal electrodes, the electrode surfaces undergo distinct redox reactions. We provide here a detailed explanation. The major difference between the toy model and the realistic model is the mechanism of charge redistribution occurring at electrode surfaces: (1) In the ideal toy model with vacuum between the pair of electrodes, when different bias potentials are applied on the pair of counter electrodes, positive charges appear at the surface of anode indicating an oxidation of Li ($Li^0 \rightarrow Li^+$), and negative charges have to emerge in the cathode side as the whole unit cell must maintain neutrality. Considering the space between electrode slabs is empty, the negative charges can only accumulate at cathode surface, which can be denoted as $Li^0 \rightarrow Li^-$. (2) Whereas in the realistic model case, we have electrolyte and SEI materials between the metal slabs with a bunch of $Li^+$ cations distributed near electrode surfaces, the anode side still undergoes Li oxidation ($Li^0 \rightarrow Li^+$), the counter negative charge, however in this case, can reduce the Li cations near the cathode surface, which can be denoted as a reduction from $Li^+ \rightarrow Li^0$. We stress again that the key difference between the ideal vacuum model and the realistic model with electrolyte in between lies in the absence vs. presence of $Li^+$ in the electrolyte environment near electrode surfaces. These $Li^+$ cations can accommodate negative balance charges in the cathode side, thus are reduced to $Li^0$ when a reducing potential is applied in a ConstP condition. We therefore would not observe $Li^0 \rightarrow Li^-$ in a realistic model case.

## 2.11 Li charge over time under different external potentials

In the main text, we validate the ConstP DP-QEq method. The supporting results for the ConstP's validations are displayed in Figure S14. Specifically, when $\phi_{\mathrm{Li},1}$ decreases from -1 to -2 V with $\phi_{\mathrm{Li},2}$ fixed at 6 V, the normalized charge of anode Li atoms (the left region with a dashed frame in **Figure 1d**) increases from ~0.31 to ~0.42 Q per Li (Q here denotes a positive unit charge), indicating more $Li^0$ atoms being oxidized to $Li^+$ ions. In contrast, when $\phi_{\mathrm{Li},2}$ increases from +6 to +7 V with $\phi_{\mathrm{Li},1}$ fixed at -2 V, the normalized charge of cathode Li atoms (the right region with a dashed frame in **Figure 1d**) decreases from ~0.12 to ~0.10 Q per Li, suggesting more $Li^+$ ions being reduced to $Li^0$ atoms.

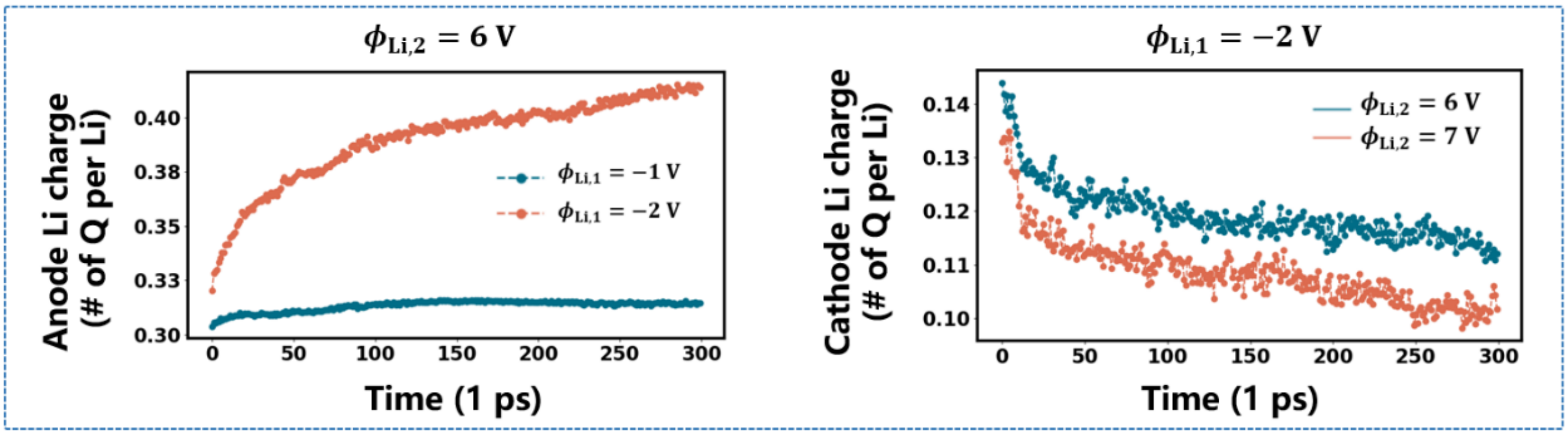

**Figure S14.** Normalized Li atomic charges over time in the anode-side (the left region marked by the dashed frame in the main text Figure 1d) and the cathode-side (the right region marked by the dashed frame in the main text Figure 1d) regions under different bias potentials (electronegativity shifts).

## 2.12 Distributions of SEI inorganic components

In Figure S15, we plot the distribution of various SEI products in the initial and final snapshots belonging to the 2nd round ConstP DP-QEq simulation. We can see that the Li aggregations first appear within the SEI inorganic component (shown in the left panel), and then inhomogeneously deposit on the cathode surface driven by the electrochemical reductive potential applied at cathode (the right panel).

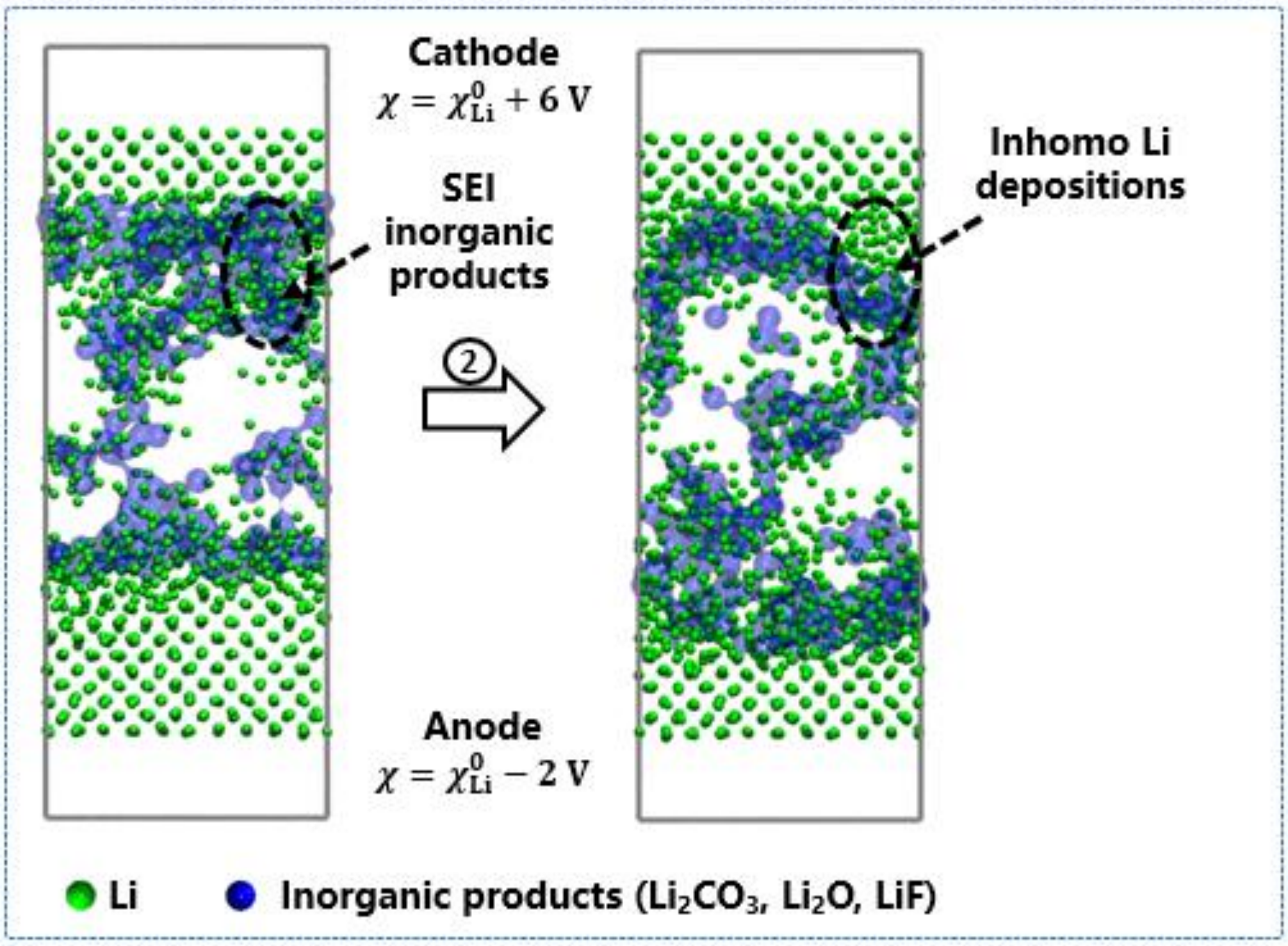

**Figure S15.** Distributions of various SEI products in the initial and final snapshots belonging to the 2nd round ConstP DP-QEq MD simulation. We only display Li atoms (green) and inorganic products (blue) for clarity.

## 2.13 ConstP testing simulations with a 3.3 Å cutoff for coordination number calculations

In all ConstP DP-QEq MD simulations, we need to decide whether to apply a biased potential on a certain Li atom based on its coordination number (CN) with a cutoff of 3.5 Å. Considering the Li-Li distance in a BCC Li metal is ~ 3.0 Å, we thus performed another series (4 rounds) of ConstP MD with a cutoff of 3.3 Å to validate the reliability of this setup in our simulation approach. We still observe a Li dendrites nucleation process in those simulations, consistent with the results reported in the main text. We display the structure plots of the testing simulations in Figure S16.

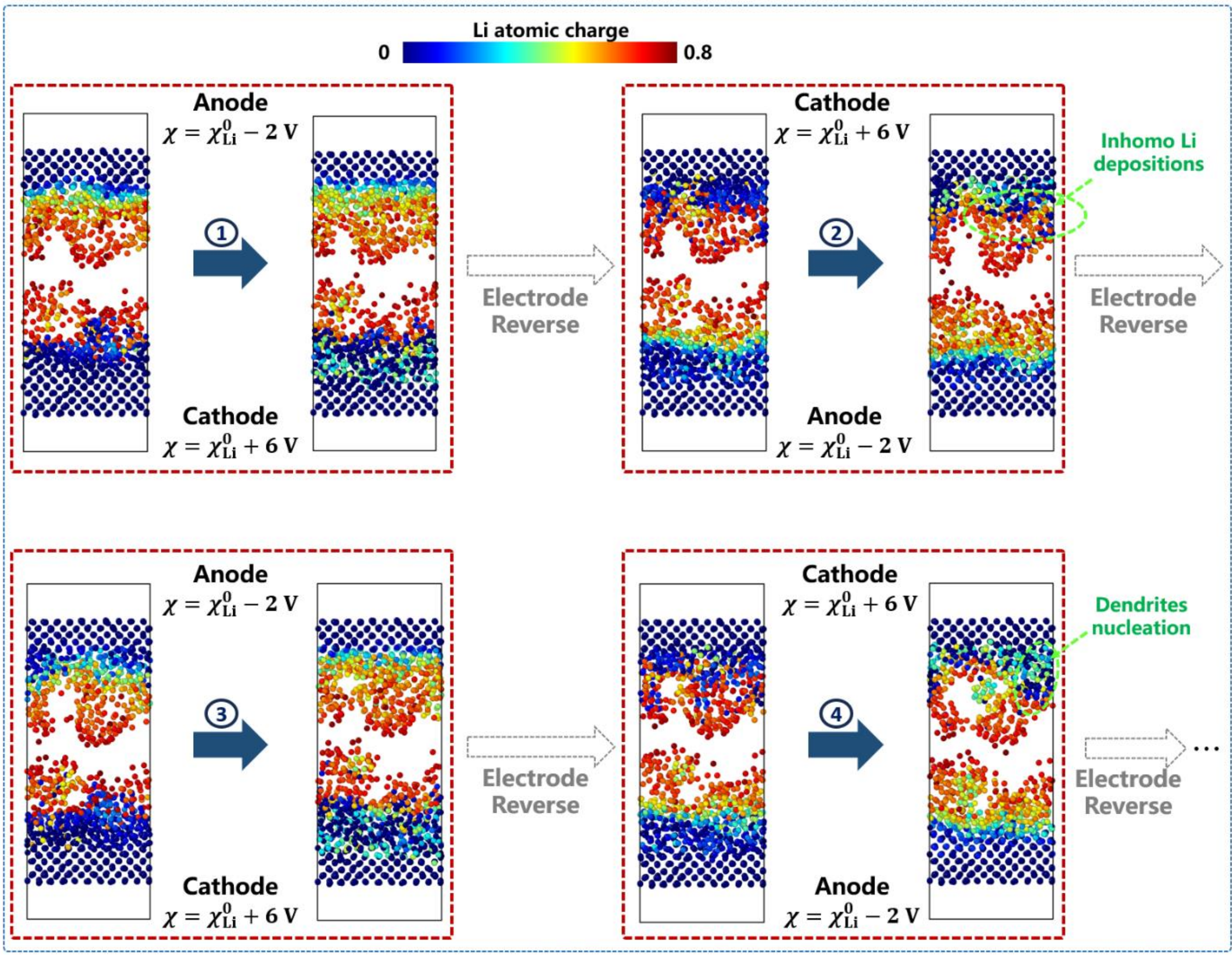

**Figure S16.** Charge state distributions of Li atoms/ions along 1- 4 rounds ConstP MD testing simulations with a cutoff of 3.3 Å which is used to decide whether to apply a bias potential on a certain Li atom. The reproduced Li dendrites nucleation region is highlighted by the green dashed circles.

## 2.14 Changes of Li-Li bond numbers between the initial and the last electrochemical cycles

We investigate the changes of normalized Li-Li bond numbers (i.e. the number of Li-Li bonds per Li atom) in the inhomogeneous Li deposition region in the last round DP-QEq ConstP simulation compared to the initial round (Figure S17). We track the same Li atoms for bond number statistics in the initial and last simulation rounds. The supporting evidences demonstrate that the Li aggregation occurs in this region, as we can see the normalized Li-Li bond number increasing from the initial to the last round.

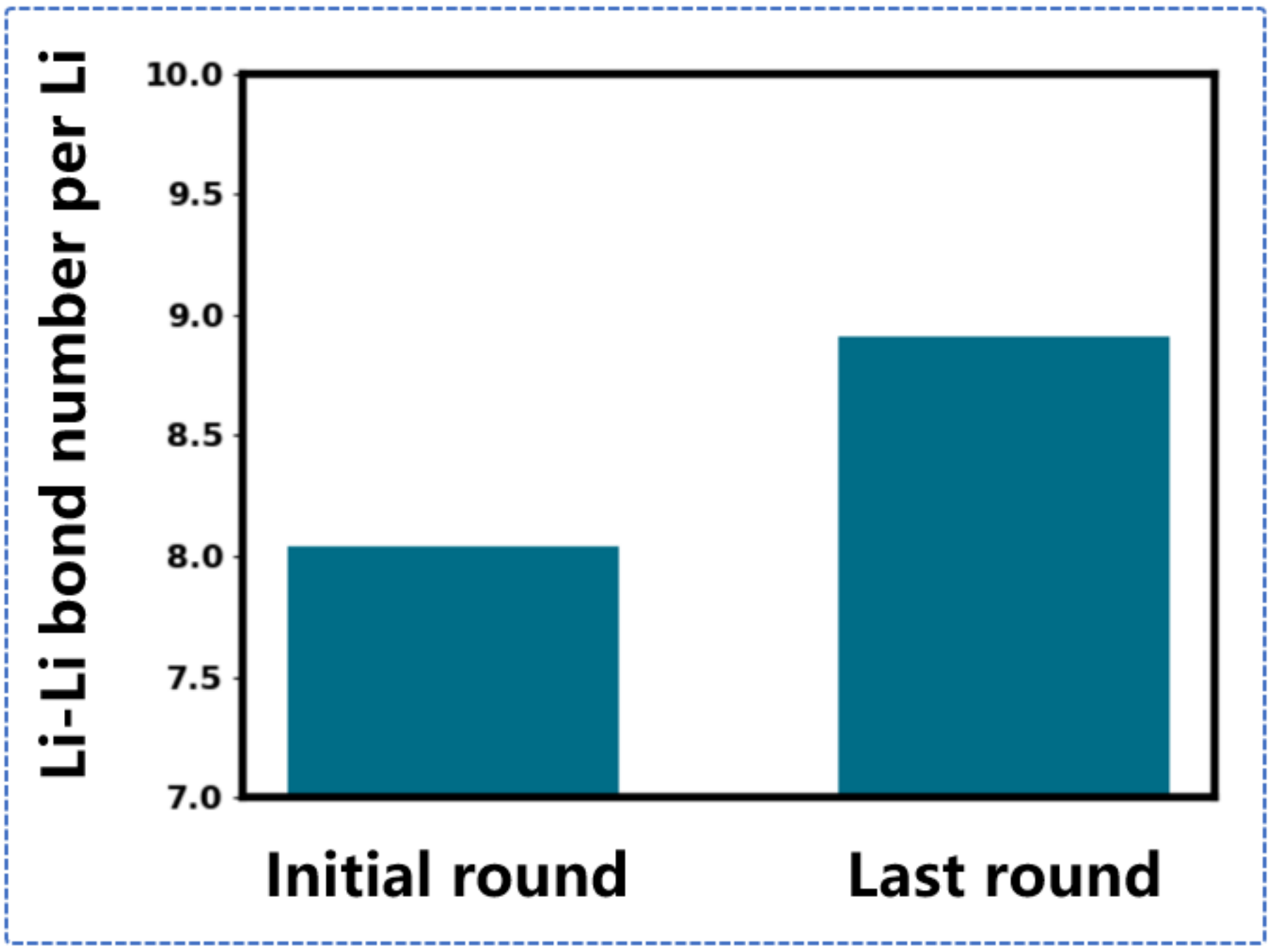

**Figure S17.** Normalized Li-Li bond numbers in the inhomogeneous Li deposition region in the initial and last rounds of ConstP MD simulations. All Li-Li pairs with distances smaller than 3.5 Å are taken into account.

## 2.15 Computational workflow of our DP-QEq MD simulations

We present the workflow of atomic model construction for DP-QEq MD simulations in Figure S18. Specifically, we first conducted a 200 ps NPT ConstQ MD simulation (pressures set as 1 bar along the *x* and *y* directions and as 100 bar along the *z* direction) for a supperlattice model to avoid cavity generations (Figure S18a → Figure S18b). We employed the Nosé-Hoover[29,30] thermostat combined with the Parrinello-Rahman[31] dynamics to apply external pressures for relaxing cell volumes in this NPT simulation. We then constructed a double-interface model containing a pair of counter electrodes by cutting from the middle of the Li metal slab in the above supperlattice model (Figure S18b → Figure S18c). We fixed the uppermost/bottommost four layers of the top/bottom Li slab to maintain a bulk environment for the interior regions of electrodes (Figure S18c → Figure S18d). The prepared double-interface model is employed in subsequent ConstQ and cyclic ConstP MD simulations. We used the 3-D periodic boundary condition throughout all simulations.

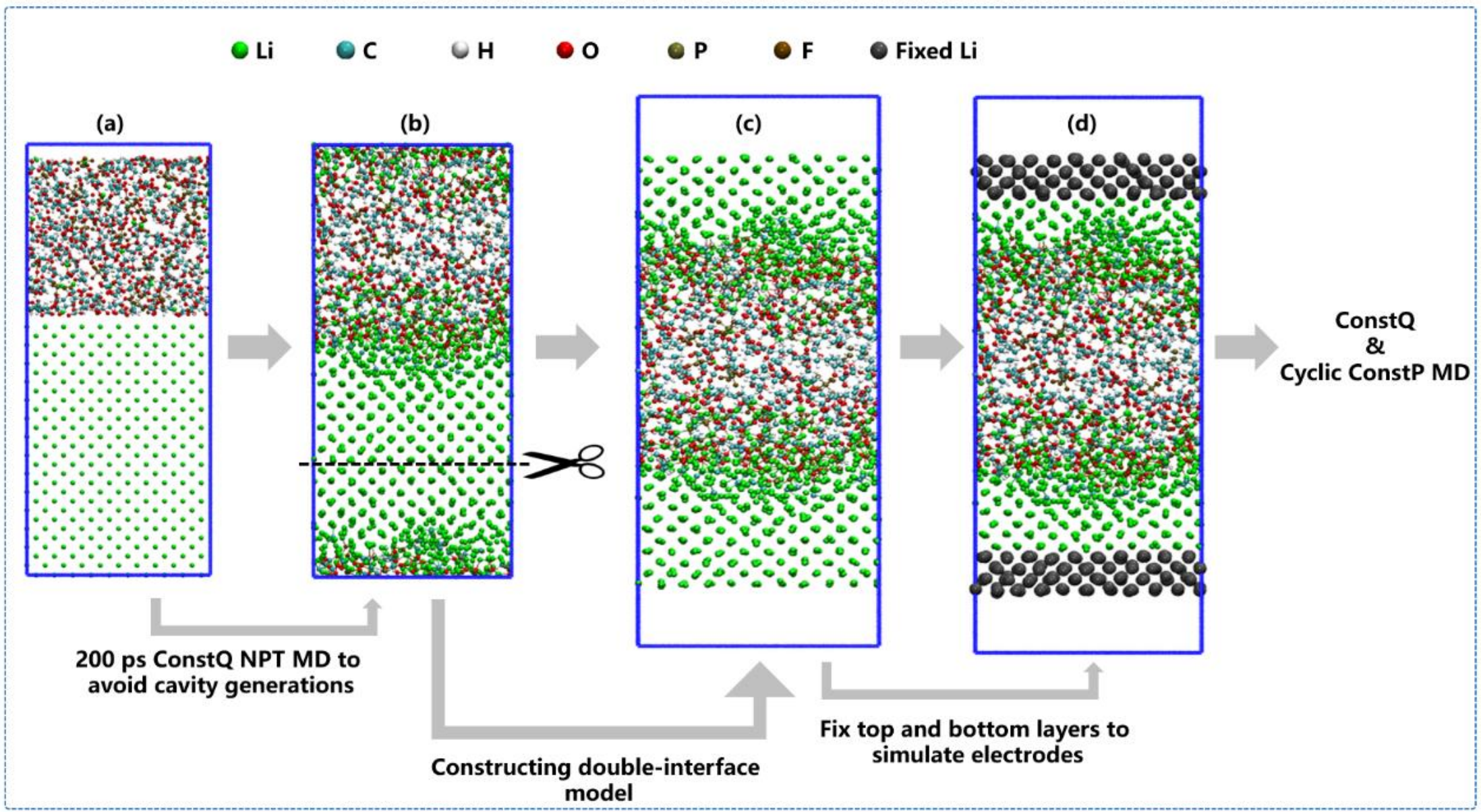

**Figure S18.** The workflow of atomic model construction for MD simulations in this study.